\documentclass[a4paper,twocolumn,showpacs,superscriptaddress,amsmath,amssymb,aps,prl,preprintnumbers]{revtex4-2}
\usepackage{bbm}
\usepackage{dsfont}
\usepackage{amsmath}
\usepackage{verbatim}
\usepackage{graphicx}
\usepackage{float}
\usepackage{braket,amsmath}
\usepackage[dvipsnames, svgnames, x11names]{xcolor}
\usepackage{array}
\usepackage{bibunits}
\usepackage{braket}
\usepackage{mathrsfs}
\usepackage{amsmath}
\usepackage{graphicx}
\usepackage{dcolumn}
\usepackage{bm}
\usepackage[capitalise]{cleveref}
\crefname{section}{Sec.}{Secs.}
\crefname{appendix}{App.}{Apps.}

\begin{document}
\emph{}

\title{Concurrent spin squeezing and light squeezing in an atomic ensemble}
\author{ Shenchao Jin}%
\affiliation{%
State Key Laboratory of Quantum Optics and Quantum Optics Devices,
Institute of Laser Spectroscopy, Shanxi University, Taiyuan, Shanxi 030006, China
}%
\affiliation{%
Collaborative Innovation Center of Extreme Optics, Shanxi University, Taiyuan, Shanxi 030006, China
}%
\affiliation{%
 Department of Physics, State Key Laboratory of Surface Physics and Key
Laboratory of Micro and Nano Photonic Structures (Ministry of Education),
Fudan University, Shanghai 200433, China
}%
\author{Junlei Duan}%
\affiliation{%
 Department of Physics, State Key Laboratory of Surface Physics and Key
Laboratory of Micro and Nano Photonic Structures (Ministry of Education),
Fudan University, Shanghai 200433, China
}
\author{Youwei Zhang}%
\affiliation{%
 Department of Physics, State Key Laboratory of Surface Physics and Key
Laboratory of Micro and Nano Photonic Structures (Ministry of Education),
Fudan University, Shanghai 200433, China
}
\author{Xichang Zhang}%
\affiliation{%
 Department of Physics, State Key Laboratory of Surface Physics and Key
Laboratory of Micro and Nano Photonic Structures (Ministry of Education),
Fudan University, Shanghai 200433, China
}
\author{\\Han Bao}%
\affiliation{%
 Department of Physics, State Key Laboratory of Surface Physics and Key
Laboratory of Micro and Nano Photonic Structures (Ministry of Education),
Fudan University, Shanghai 200433, China
}%
\affiliation{QUANTUM, Johannes Gutenberg-Universit\"{a}t Mainz, 55128 Mainz, Germany}
\author{Heng Shen}%
 \affiliation{%
State Key Laboratory of Quantum Optics and Quantum Optics Devices,
Institute of Opto-electronics, Shanxi University, Taiyuan, Shanxi 030006, China
}%
\affiliation{%
Collaborative Innovation Center of Extreme Optics, Shanxi University, Taiyuan, Shanxi 030006, China
}%
\author{Liantuan Xiao}%
\affiliation{%
State Key Laboratory of Quantum Optics and Quantum Optics Devices,
Institute of Laser Spectroscopy, Shanxi University, Taiyuan, Shanxi 030006, China
}%
\affiliation{%
Collaborative Innovation Center of Extreme Optics, Shanxi University, Taiyuan, Shanxi 030006, China
}%
\author{Suotang Jia}%
\affiliation{%
State Key Laboratory of Quantum Optics and Quantum Optics Devices,
Institute of Laser Spectroscopy, Shanxi University, Taiyuan, Shanxi 030006, China
}%
\affiliation{%
Collaborative Innovation Center of Extreme Optics, Shanxi University, Taiyuan, Shanxi 030006, China
}%
\author{ Mingfeng Wang}%
 \email{mfwang@wzu.edu.cn}
\affiliation{%
 Department of Physics, Wenzhou University, Zhejiang 325035, China
}%
\author{Yanhong Xiao}
\email{yxiao@fudan.edu.cn}
\affiliation{%
State Key Laboratory of Quantum Optics and Quantum Optics Devices,
Institute of Laser Spectroscopy, Shanxi University, Taiyuan, Shanxi 030006, China
}%
\affiliation{%
Collaborative Innovation Center of Extreme Optics, Shanxi University, Taiyuan, Shanxi 030006, China
}%
\affiliation{%
Department of Physics, State Key Laboratory of Surface Physics and Key
Laboratory of Micro and Nano Photonic Structures (Ministry of Education),
Fudan University, Shanghai 200433, China
}%


\begin{abstract}
Squeezed spin states and squeezed light are both key resources for quantum metrology and quantum information science, but have been separately investigated in experiments so far. Simultaneous generation of these two types of quantum states in one experiment setup is intriguing but remains a challenging goal. Here we propose a novel protocol based on judiciously engineered symmetric atom-light interaction, and report proof-of-principle experimental results of concurrent spin squeezing of $0.61\pm0.09~\mathrm{dB}$ and light squeezing of $0.65^{+0.11}_{-0.10}~\mathrm{dB}$ in a hot atomic ensemble. The squeezing process is deterministic, yielding fixed squeezing directions for both the light field and the collective atomic spin. Furthermore, the squeezed light modes lie in the multiple frequency sidebands of a single spatial mode. This new type of dual squeezed state is applicable for quantum enhanced metrology and quantum networks. Our method can be extended to other quantum platforms such as optomechanics, cold atom and trapped ions.
\end{abstract}
\maketitle

Entanglement states such as squeezed spin states (SSS) and squeezed light~\cite{messikh_spin_2003, korbicz_spin_2005, toth_spin_2009} have been under active investigations due to their relevance for precision measurements~\cite{PhysRevLett.96.010401,NP2,RevModPhys.90.035005}, such as light interferometers \cite{prlll} and spin sensors~\cite{RevModPhys.74.1153,PhysRevLett.105.053601,PhysRevLett.127.193601,PhysRevLett.104.093602,PhysRevA.86.023803, VuleticNature2020, Malia2022}. However, the generation and utilization of light squeezing and spin squeezing have largely been separate in different setups. The flying spin (light) and stationary spin (atoms) often involve distinct schemes~\cite{RevModPhys.90.035005} for squeezing although their squeezed states both involve pairs of correlated excitations~\cite{PhysRevLett.57.2520,PhysRevLett.109.173603, PhysRevA.47.5138, PhysRevA.66.043815}. 
Studying the analogies and differences between atoms and light in the context of quantum squeezing is an intriguing topic, with other studies in view of wave-particle duality and mathematical description of their coherent states~\cite{PhysRevA.6.2211}. Practically, concurrent squeezing of the two can save resources and have applications in quantum metrology and quantum information science. 

Despite a few theory proposals~\cite{PhysRevLett.97.143602, PhysRevA.107.033517}, concurrent light squeezing and atomic spin squeezing remains an experimental challenge because the two squeezing requires different physics processes in general. Light squeezing is usually based on nonlinear atom-light interactions (such as wave mixing \cite{PhysRevLett.55.2409} or parametric processes \cite{Wu:87}) and leaves atomic states nearly intact, whereas in light-mediated spin squeezing, light serves as the agent in correlating the spin while the spin often does not provide the necessary feedback mechanism for light squeezing~\cite{PhysRevLett.105.193602, PhysRevA.96.013823, NP23}. A generic form of Hamiltonian $\hat H = \mu_- \hat b ^\dagger \hat a ^\dagger + \mu_+ \hat b ^\dagger \hat a + \mathrm{h.c.}$ \cite{Nature.s41586-019-1320-2,science.aac5138,kubasik_polarization-based_2009, dalla_torre_dissipative_2013,2022arXiv220800024M} has been studied, where light ($\hat a$) and spin ($\hat b$) act as quantum bus for one another and in principle allows concurrent squeezing. In practice though, the spin oscillator often has finite frequency $\Omega$ and the additional term $\Omega \hat b ^\dagger \hat b$ induces quantum back-action noise by rotating and mixing the two orthogonal quantum quadratures, which can spoil spin squeezing. Up to now, Hamiltonians of such type in atomic and optomechanical systems have led to either entangled spin \cite{PhysRevLett.107.080503}, or entangled/squeezed light \cite{Nature.s41586-019-1320-2,Wasilewski:09}, but not both.

\begin{figure*}[htbp]
	\centering
	\includegraphics[width=2\columnwidth]{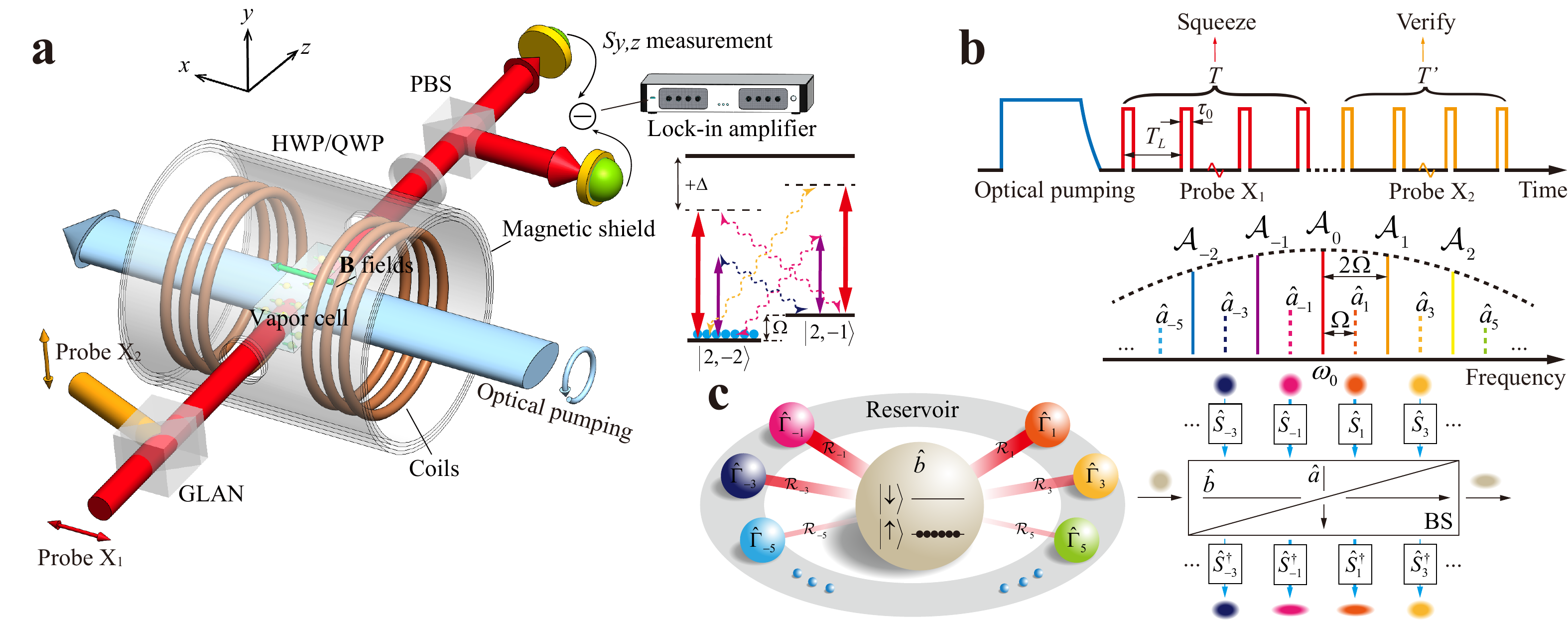}
	\caption{
(Color online) (a) Experimental schematics.
The paraffin-coated rubidium vapor cell is located in a four-layer magnetic shield, with a Zeeman splitting of $\Omega = 2\pi\times 499.60~\mathrm{kHz}$ from a bias magnetic field.
The optical pumping light travels along $x$ axis.
Both $\rm{X}_1$ and X$_2$ probe lasers propagate along the $z$-axis, and their Stokes components $S_z$ and $S_y$ will be respectively demodulated at the Larmor frequency and recorded by a lock-in amplifier. HWP, half-wave plate, QWP, quarter-wave plate, PBS, polarization beam splitter, GLAN,  Glan-laser calcite polarizer.
Right: level diagram of the probe X$_1$ with stroboscopic driving frequency $\omega_m = 2\Omega$.
The red solid arrows are the carrier (zeroth sideband) of probe X$_1$ with frequency $\omega_0$ while the purple solid arrows denote the first sideband with frequency $\omega_0-2\Omega$.
The dashed arrows represent the Stokes and anti-Stokes quantum fields in the $y$-polarization (viewed along $z$-axis).
(b) Time-domain representation of pulse sequence (top) and the corresponding spectrum of the probe pulses (bottom).
Before interaction, the optical pumping pulse prepares the atoms to CSS.
Then the stroboscopic probe X$_1$ pulse creates spin squeezing of the atomic ensemble, while the output X$_1$ pulse itself is squeezed at the same time.
An extra stroboscopic probe X$_2$ pulse then verifies the generated SSS.
All the three pulses have slowly varying rising and falling edges \cite{PRJ.413288} to minimize excess noises.
(c) Pictorial representation of the atom-light interaction. It can be understood as either (left) an atomic mode interacting with a squeezed reservoir or (right) an atom-light beam splitter (BS) evolution sandwiched between the sideband squeezing evolutions (\emph{see text}).}
\label{fig1}
\end{figure*}

Here, we propose a novel protocol enabling squeezing of both light and a collective spin oscillator, and report proof-of-principle experiments. By using stroboscopic atom-light interaction with periodic short optical pulses of coherent state, the frequency mismatch between light and atoms is compensated as in Floquet engineering \cite{geier_floquet_2021,jiang_floquet_2022}. This allows generation of a comb-like squeezed vacuum light state which then transfers to atoms producing spin squeezing. Previously, stroboscopy was employed in back-action evading
measurements and conditional spin squeezing \cite{NP23, Nature.581.7807}. We experimentally demonstrate deterministic and unconditional spin squeezing, with fixed squeezing directions for both light and spin. 

The principle of our scheme for concurrent squeezing is illustrated in Fig.~\ref{fig1}.
A $^{87} \mathrm{Rb}$ atomic ensemble with $N_A>10^{11}$ atoms are prepared in the ground state sublevel $\ket{\uparrow}\equiv 5^2\mathrm{S}_{1/2}\ket{F=2,m_F=-2}$ to produce a CSS along a bias magnetic field \cite{PhysRevA.63.055601} in the $x$-direction, which can be treated as a two-level system with the neighboring sublevel $\ket{\downarrow}\equiv\ket{F=2,m_F=-1}$.
The creation of the atomic excitation can be described by the collective operator $\hat{b}^\dag  = \sum_{i = 1}^{N_A} \ket{\downarrow}_i\bra{\uparrow} /\sqrt{N_A}$, satisfying the commutation relation $[\hat{b},\hat{b}^\dag]=1$. A train of ultrashort and relatively strong linearly-$x$-polarized light pulse with central frequency $\omega_0$ is sent through the sample to experience a near-resonant interaction, producing Raman scattered and collectively enhanced quantum field $\hat{a}$ in the $y$-polarization.
These square-wave pulses have a width $\tau_0$ and are  $T_L=\pi/\Omega$ away from each other in the time domain to ensure that each pulse enters the ensemble encountering exactly either the $+z$ or the $-z$ component of the atomic angular momentum (which undergoes Larmor precession around the $B$-field), as shown in \cref{fig1}(b). With such setting, the periodically changing temporal profile function of the incident light in the frequency domain is: $\phi(t)=\sum_{n=-\infty}^\infty \mathcal{A}_n e^{i2n\Omega t}$, with the Fourier coefficients $\mathcal{A}_n=d~\mathrm{sinc}(\pi n d)$ and the duty cycle $d = \tau_0/T_L$. Obviously, this spectral distribution is a frequency comb \cite{lightcomb} with central frequency $\omega_0$, overall width $\propto 1/d$, and comb-tooth separation $2\Omega$ [\cref{fig1}(b)].
The system can be described by canonical variables under the Holstein-Primakoff approximation \cite{PhysRev.58.1098}, $\hat{x}_A = - \hat{J}_y/\sqrt{|J_x|} = (\hat{b}+\hat{b}^\dag)/\sqrt{2}, \hat{p}_A = \hat{J}_z/\sqrt{|J_x|} = -i(\hat{b}-\hat{b}^\dag)/\sqrt{2}$ for the collective spin and $\hat{x}_L = \hat{S}_y/\sqrt{|S_x|} = (\hat{a}+\hat{a}^\dag)/\sqrt{2}, \hat{p}_L = \hat{S}_z/\sqrt{|S_x|} = -i(\hat{a}-\hat{a}^\dag)/\sqrt{2}$ for light, where $\hat{J}_i$ are collective angular momenta of the atomic ensemble and $\hat{S}_i$ are Stokes operators for light with $i=x,y,z$, and the minus sign before $\hat{J}_y$ comes from the atomic polarization direction (see section 1 of the Supplemental Material \cite{supplemental}).

In the weak excitation limit, one may adiabatically eliminate the atomic exited states and obtain the effective interaction Hamiltonian (section 1 of~\cite{supplemental}) of the type:
\begin{eqnarray}\label{eq1}
{{\hat H_{\rm{int}}}} &\propto& {{\hat b}^\dag }\sum\limits_{k =  - \infty }^\infty  {\mathcal{R}_k\left( {{{\hat a}_{k}}\cosh r_k - \hat a_{k}^\dag \sinh r_k} \right)}  + {\rm{h}}{\rm{.c}}{\rm{.}},
\label{eq1}
\end{eqnarray}
where the integer $|k|$ is odd and $\hat{a}_{k}^\dag$ denotes creation operator of the sideband mode with frequency $\omega_0+k\Omega$. The sideband-dependent coupling constant ${\mathcal{R}_k} = \sqrt{\mathcal{C}_{+}\mathcal{C}_-}$ and the parameter ${r_k} = \ln (\mathcal{C}_{+}/\mathcal{C}_-)$ with ${\cal C}_{\pm}={{\cal A}_{\frac{1}{2}(k - 1)}}{\mu _ + } \pm {\rm{ }}{{\cal A}_{\frac{1}{2}(k + 1)}}{\mu _ - }$, where ${\mu _ \pm } =  \kappa (1 \pm {\zeta ^2})/2\sqrt{L}$ describe Stokes and anti-Stokes scattering rates, with total interaction length $L$, coupling strength $\kappa$ (proportional to the square root of $N_A$ times the incident photon number), and $\zeta^2$ proportional to the ratio of the coefficients of the vector and tensor parts of the atomic polarizability. This interaction has a simple and interesting form. It describes a beamsplitter (BS) type interaction between the atomic mode $\hat b$ and the photonic Bogoliubov mode $\hat \Gamma_{k}$ ($\equiv\hat S^\dag_{k} \hat a_{k}\hat S_{k}={{{\hat a}_{k}}\cosh r_k - \hat a_{k}^\dag \sinh r_k}$), where the squeeze operator ${\hat S_{k}} = \exp [ {r_k}(\hat a_{k}^2 - \hat a_{k}^{\dag 2})/2]$. The Bogoliubov modes $\{\hat \Gamma_k\}$ bears close analogy to the \emph{squeezed vacuum reservoir} \cite{scully_quantum_1997, PhysRevA.41.3782}, passing squeezing to atoms as in Eq.~(\ref{eq1}) and illustrated in Fig.~1(c). More specifically, the evolution of such interaction is equivalent to $\hat S_k^\dag \hat b^\dag \hat a_{k}\hat S_{k} + \mathrm{h.c.}$, i.e., an atom-light BS evolution sandwiched between the sideband squeezing evolutions $\{\hat S_k\}$ and $\{\hat S_k^\dag\}$, as shown in \cref{fig1} (c). The vacuum sideband modes are first squeezed by $\hat S_k$ (along $\hat x_L$) and then the squeezing properties are transferred from light to atoms via the BS evolution, which returns light to coherent state but yields spin squeezing along $\hat p_A$, and finally the light modes are squeezed via $\hat S^{\dag}_k$ along $\hat p_L$. This explains why both light and spin are squeezed in the $\hat p$ quadrature, i.e., along $S_z$ and $J_z$ respectively. 

The propagation and evolution equations of the canonical variables can be derived (section 1 of ~\cite{supplemental}):
\begin{subequations} \label{eq2}
\begin{eqnarray}
\frac{\partial {{\hat x}_L}}{{\partial z}}  =& \frac{\kappa}{\sqrt{L}} \phi {{\tilde {\hat p}}_A},
&\frac{{\partial {{\tilde {\hat x}}_A}}}{{\partial t}} = \Omega {{\tilde {\hat p}}_A} + \frac{\kappa}{\sqrt{L}} \phi {{\hat p}_L},\\
\frac{\partial {{\hat p}_L}}{{\partial z}}  =& -\frac{{\zeta ^2}\kappa}{\sqrt{L}} \phi {{\tilde {\hat x}}_A},
&\frac{{\partial {{\tilde {\hat p}}_A}}}{{\partial t}} =  - \Omega {{\tilde {\hat x}}_A} - \frac{{\zeta ^2}\kappa}{\sqrt{L}} \phi {{\hat x}_L}.
\end{eqnarray}
\end{subequations}
Here ${\tilde {\hat x}}_A$ and $ {\tilde {\hat p}}_A$ denote the quadratures for an atomic slice around a given $z$.
These equations describe the information exchange between light and atoms.
One can clearly see that the flying photons evolve along the propagation distance, while the residing atoms evolve with time, showing the difference between light and spins. The motivation for stroboscopy can be also seen here: squeezing in $\hat p_A$ is mediated by $\hat x_L$, but the $\Omega$ rotation term brings anti-squeezing noise in $\hat x_A$, which must be eliminated by stroboscopy in $\phi$ as defined before. In the frequency domain, frequency sidebands due to stroboscopy enables the excitation of correlated spin pairs responsible for spin squeezing, as illustrated in Fig.~1(a), via the two Raman processes containing the generated quantum light (two violet wavy arrows), the input light's carrier (red straight arrow) and first order sideband (violet straight arrow).   

 \begin{figure}[htbp]
\centering
\includegraphics[width=1.05\columnwidth]{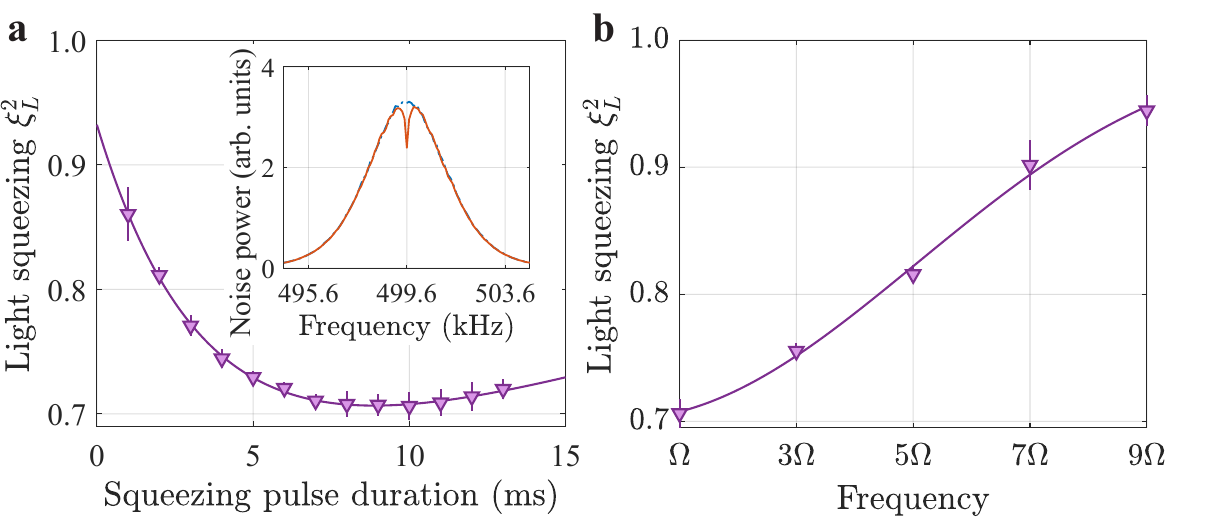}
\caption{(a) Light squeezing vs. squeezing time $T$, with mean power $1.18~\mathrm{mW}$ and duty cycle $d=0.08$.
Inset: exemplary squeezed light spectrum of the light X$_1$ around $\omega=\Omega$.
The red solid curve denotes the power spectrum of the homodyne signal $\hat p_L$, while the blue dash-dotted line represents the measured shot noise reference.
The noise spectrum is averaged from $10,000$ runs of the pulse sequence as shown in Fig.~\ref{fig1}(b).
(b) Light squeezing at different frequencies.
For each data point, the demodulation frequency of the lock-in is set to the corresponding $x$-axis value.
In the inset and (b), $T = 10~\mathrm{ms}$.
In (a) and (b), the triangle with error bar is the experimental data with one standard deviation from 5 identical experiments (each with $10,000$ runs of the pulse sequence), and the solid curves are the model fits (section 1 of~\cite{supplemental}).
The SN limit is 1.0.
}
\label{fig2}
\end{figure}

Our experiment [\cref{fig1}(a)] utilizes a hot atomic vapor cell with paraffin coating \cite{Paraffin}.
The atoms are first optically pumped to $\ket{\uparrow}$, with a measured degree of polarization about $97.9\%$ which has $6\%$ extra noise in spin variance compared to an ideal CSS of $100\%$ polarization.
Then, a stroboscopic laser pulse X$_1$ is sent through the atoms to induce squeezing for both the spin and the light.
Light squeezing is verified by measuring X$_1$'s Stokes component $S_z$ ($\hat{p}_L$ quadrature).
To verify spin squeezing, another stroboscopic and far-detuned laser pulse X$_2$ is applied to create a standard back-action evading QND interaction $\propto \hat{p}_L\hat{p}_A$ \cite{Nature.581.7807}, where $\hat{p}_A$ can be read out by detecting X$_2$'s $S_y$ ($\hat{x}_L$) component (section 2 of ~\cite{supplemental}).

We first characterize the multi-mode light squeezing.
The balanced homodyne $S_z$ measurement on X$_1$ yields the $\hat p_{L}$ signal, with a typical power spectrum shown in the inset of \cref{fig2}(a).
From the noise spectrum, one can see that there exists an apparent dip at the Larmor frequency $\omega=\Omega$, indicating light squeezing.
A careful calibration from the spectrum normalized to that of the photon shot noise gives the degree of squeezing in Fig.~\ref{fig2}(a), showing that squeezing increases first with the length of the X$_1$ pulse because of lager coupling strength, and then slowly degrades due to the decay of the atomic macroscopic spin (section 2 of~\cite{supplemental}).
The optimal light squeezing is $-10~\mathrm{log}_{10}(\xi_{L}^2) = 1.51\pm0.07~\mathrm{dB}$ with squeezing pulse duration $T = 10~\mathrm{ms}$.
By using different demodulation frequencies, we can obtain squeezing at frequencies $m\Omega$ ($m=1,3,5,7,...$) [\cref{fig2}(b)], showing that the amount of squeezing decreases for higher frequencies, following a trend consistent with our theoretical prediction (section 1 of~\cite{supplemental}).
The multi-sideband squeezing here is closely related to a multipartite entanglement state \cite{Flammia_2009} applicable for quantum information processing~\cite{RevModPhys.77.513}.

\begin{figure}[tp]
\centering
\includegraphics[width=1.05\columnwidth]{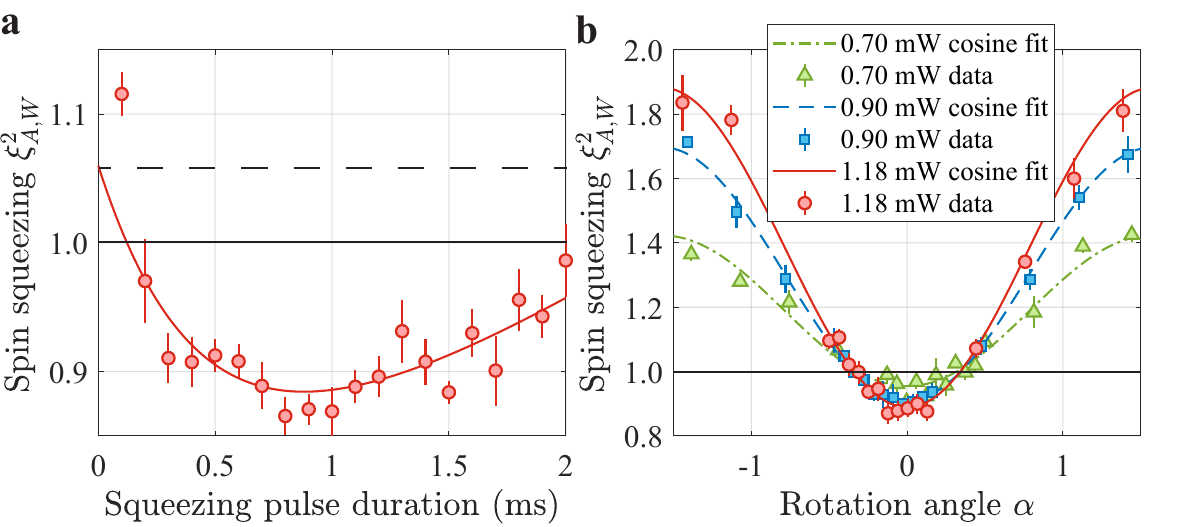}
\caption{ (a) Spin squeezing vs. squeezing time $T$, with mean power $1.18~\mathrm{mW}$ and duty cycle $0.08$.
The circle with error bar is the experimental data extracted from the pulse X$_2$ detection with decaying time mode $f(T') = e^{-\gamma' T'}, \gamma'^{-1} = 1.3~\mathrm{ms}$, mean power $1.24~\mathrm{mW}$ and duty cycle $0.1$.
The line is fixed to 1.06 at $T=0$ (shown by the black dashed line) because of the $97.9\%$ spin polarization. The first data point deviates from the fitted curve due to excess technical noise originating from an overly short and sharp pulse violating the adiabaticity condition~\cite{PRJ.413288}.
(b) Spin squeezing for different spin quadrature directions in the $y$-$z$ plane, with $T = 1.0~\mathrm{ms}$, mean power $1.18~\mathrm{mW}$ and duty cycle $0.08$.
The rotation angle $\alpha$ is the angle relative to $\hat{p}_A$.
In (a) and (b), the solid lines are the model fits (section 1 of~\cite{supplemental}), and
the error bars (one standard deviation) are derived from 5 identical experiments, each including $10,000$ repetitions of the pulse sequence shown in Fig.~\ref{fig1}(b).
The black line at 1.0 is the PNL.
}
\label{fig3}
\end{figure}

Then we proceed to spin squeezing evaluated by Wineland criterion \cite{PhysRevA.50.67} $\xi_{A,W}^2 = e^{2T/T_1}  \mathrm{Var}(\hat{p}_{A})_\mathrm{SSS} /  \mathrm{Var}(\hat{p}_{A})_\mathrm{PNL}$, where $\mathrm{Var}(\hat{p}) = \langle \hat{p}^2 \rangle - \langle \hat{p} \rangle^2$ and the prefactor $e^{2T/T_1}$ denotes the effect of the macroscopic-spin decay with relaxation time $T_1 = 18~\mathrm{ms}$.
The projection noise limit (PNL) $\mathrm{Var}(\hat{p}_{A})_\mathrm{PNL}$ is equal to $4/5$ times the measured noise of the thermal state \cite{NP23, Nature.581.7807}.
\cref{fig3}(a) presents the measured time-evolution of spin squeezing, which agrees well with the theoretical result, as shown by the solid line fit.
The competition between the coherent interaction and decoherence leads to an optimal spin squeezing $-10~\mathrm{log}_{10}(\xi_{A,W}^2)=0.63\pm0.08~\mathrm{dB}$ at $T = 0.8~\mathrm{ms}$, which has suffered a degradation of about $0.39~\mathrm{dB}$ due to the main spin vector shortening induced by spontaneous emission and other decoherences during squeezing.

To examine the spin squeezing direction, we measure the spin quadrature along a direction $\alpha$: $\hat{q}_\alpha = \hat{p}_A \cos (\alpha/2) + \hat{x}_A \sin (\alpha/2)$, where $\alpha$ is varied by tuning the time interval between the X$_1$ and X$_2$ pulse.
During this interval, the transverse spin ellipse undergoes Larmor precession for a fraction of a Larmor period and the spin decoherence is negligible.
The spin direction is experimentally calibrated by radio frequency signal excitation.
\cref{fig3}(b) plots the amount of squeezing in different directions, indicating that the squeezing direction is along $\hat{p}_A$ even for different interaction strength.
This proves that our method produces deterministic SSS with fixed squeezing direction. In fact, The interaction (\ref{eq1}) always squeezes the spin component parallel to the propagation direction of the light and broadens the orthogonal component at the same time.
In contrast to SSS produced by QND schemes \cite{Kuzmich_1998,PhysRevLett.85.1594} that dependends on
the measurement result and SSS via one-axis twisting \cite{PhysRevA.47.5138} whose squeezing direction varies with the interaction strength, SSS with fixed squeezing direction may be convenient for use in quantum metrology and quantum information applications \cite{Malia2022, PhysRevLett.104.133601, PhysRevLett.125.210503,RevModPhys.77.513}.

\begin{figure}[b]
\centering
\includegraphics[width=1.05\columnwidth]{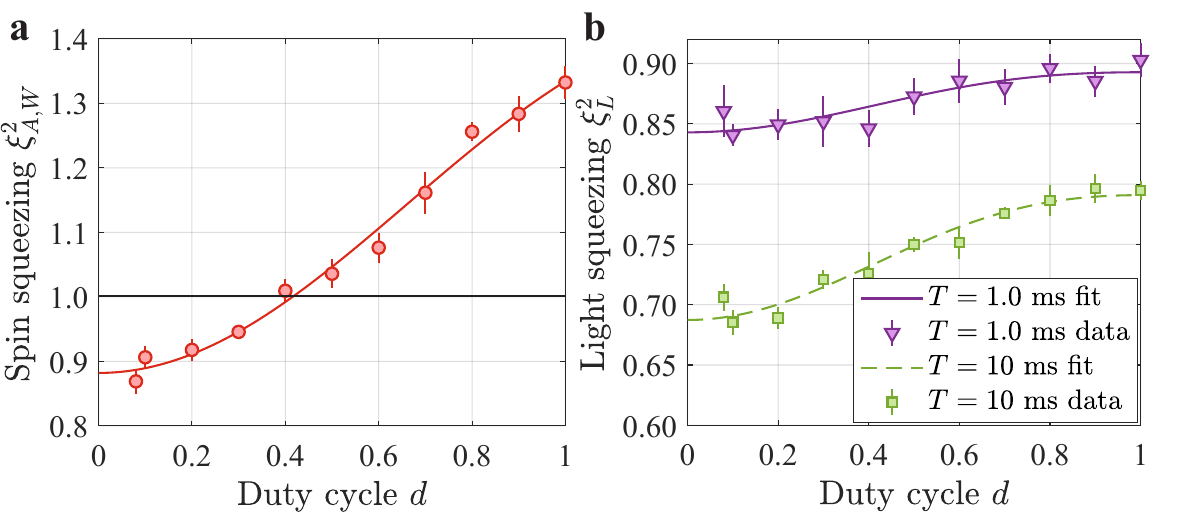}
\caption{ (a) Spin squeezing and (b) light squeezing (measured at $\omega=\Omega$) vs. duty cycle.
Red circles, purple triangles and green squares are experimental data.
The curves are the model fits (section 1 of~\cite{supplemental}) with function $b_1 + b_2~ \mathrm{sinc}(\pi d)$ for (a) and function $c_1 + c_2~\mathrm{sinc}^2(\pi d)$ for (b).
The mean power of X$_1$ is kept at $1.18~\mathrm{mW}$ in (a) and (b), while X$_2$ has a mean power $1.24~\mathrm{mW}$ with duty cycle $0.1$ in (a).
$T = 1.0~\mathrm{ms}$ in (a).
The error bars (one standard deviation) are derived from 5 identical experiments, each including $10,000$ repetitions of the pulse sequence shown in Fig.~\ref{fig1}(b).
The black line at 1.0 indicates the PNL in (a) and SN limit (not shown) in (b).
}
\label{fig4}
\end{figure}

Next, we present the result of concurrent spin and light squeezing and discuss their different behaviors. As shown in \cref{fig4}, with the same parameter $T = 1.0~\mathrm{ms}$, concurrent squeezing can be observed for $d$ smaller than 0.4, and when $d = 0.08$ we obtain concurrent spin squeezing of $0.61\pm0.09~\mathrm{dB}$ and light squeezing of $0.65^{+0.11}_{-0.10}~\mathrm{dB}$.
The measured squeezing for both the spin and light follows a similar trend. However, for large $d$, spin squeezing disappears, while light squeezing still exists even when $d=1$.
Therefore, stroboscopy is necessary for spin squeezing here but is only slightly beneficial for light squeezing.
We note that, although the squeezing process for light mode $\hat a$ and spin mode $\hat b$ is similar as indicated by the Hamiltonian $\propto \mu_- \hat b ^\dagger \hat a ^\dagger + \mu_+ \hat b ^\dagger \hat a + \mathrm{h.c.}$, their major difference is that, as shown by the propagation and evolution equations [Eq.~(\ref{eq2})], atoms reside in the vapor cell and the atom-light interaction effects accumulate with time, rendering them more sensitive to noises especially the residual quantum back-action noises, whereas the light field always enters the atomic medium fresh as coherent state. Another distinction is that, the collective atomic spin here is a single frequency harmonic oscillator, while the light pulse in free space is strictly speaking multimode in frequency~\cite{PhysRevA.83.052312}. Consequently, for light, it is possible to use the atomic mode as a mediator to establish entanglement between different light modes (for instance the two first order sidebands), which is the origin of light squeezing in absence of stroboscopy ($d=1$); for atoms, there is only single-mode squeezing where stroboscopy of light is indispensible for frequency match in spin pair excitations.
 
Finally, we note that the moderate squeezing in our experiment is mostly due to the relatively small optical depth $\alpha_0$, large spin decoherence rate and light losses. For a given $\alpha_0$, the optimal spin squeezing follows $\xi _{A,W}^2 \propto 2(\sqrt{1+\alpha_0}-1)/\alpha_0$, predicting higher squeezing for larger optical depth (section 2 of~\cite{supplemental}), while light squeezing follows a similar trend. Thus, improved squeezing in both atoms and light is expected by applying optical cavities \cite{NP23}, longer vapor cells~\cite{PhysRevLett.130.203602}, multi-pass schemes \cite{PhysRevLett.94.023003, PhysRevLett.105.193602}, or by enhanced quality in the antireflection coating as well as in the wall coatings~\cite{Alkene} including those enduring higher temperatures~\cite{OTS}. 

Our protocol can be applied in other quantum systems such as optomechanical systems \cite{Nature.s41586-019-1320-2,science.aac5138}, cold atoms \cite{kubasik_polarization-based_2009, dalla_torre_dissipative_2013}, trapped ions \cite{gilmore_quantum-enhanced_2021}, and superconducting circuits~\cite{2022arXiv220800024M} etc., where similar Hamiltonian can be engineered. 
Compared to separately prepared independent spin squeezing and light squeezing, concurrent squeezing is beneficial for various applications thanks to the spectral and bandwidth matching between the atoms and the squeezed light. For example, as we propose (detailed in section 3 of \cite{supplemental}), in quantum metrology, one could first use the squeezed spin for sensing \cite{PhysRevLett.104.133601} and then use the concurrently produced squeezed light for efficient readout of the atomic state \cite{PhysRevA.73.022331, PhysRevLett.115.093604}, which enhances the overall sensitivity; in quantum state preparation, reflecting the squeezed light back into the squeezed atoms again could further increase spin squeezing; in quantum networks \cite{kimble_quantum_2008}, spin squeezed atomic ensembles can serve as low noise quantum nodes \cite{fernholz_spin_2008}, while the simultaneously squeezed light can be utilized to connect and entangle these nodes \cite{muschik_efficient_2006, muschik_efficient_2006, schrader_neutral_2004}, where spin squeezing can boost the entanglement generation efficiency.

This work is supported by the Natural Science Foundation
of China (NSFC) under Grants No.~12161141018, No.~12027806, No.~12222409, Fund for Shanxi ``1331 Project", and National Key R\&D Program of China under Grant No. 2020YFA0309400.

\bibliography{ref} 

\end{document}



\title{Supplemental material for Concurrent spin squeezing and light squeezing in an atomic ensemble}
\author{ Shenchao Jin}%
\affiliation{%
State Key Laboratory of Quantum Optics and Quantum Optics Devices,
Institute of Laser Spectroscopy, Shanxi University, Taiyuan, Shanxi,
030006, China
}%
\affiliation{%
Collaborative Innovation Center of Extreme Optics, Shanxi University, Taiyuan 030006, China
}%
\affiliation{%
 Department of Physics, State Key Laboratory of Surface Physics and Key
Laboratory of Micro and Nano Photonic Structures (Ministry of Education),
Fudan University, Shanghai, 200433, China
}%
\author{Junlei Duan}%
\affiliation{%
 Department of Physics, State Key Laboratory of Surface Physics and Key
Laboratory of Micro and Nano Photonic Structures (Ministry of Education),
Fudan University, Shanghai, 200433, China
}
\author{Youwei Zhang}%
\affiliation{%
 Department of Physics, State Key Laboratory of Surface Physics and Key
Laboratory of Micro and Nano Photonic Structures (Ministry of Education),
Fudan University, Shanghai, 200433, China
}
\author{Xichang Zhang}%
\affiliation{%
 Department of Physics, State Key Laboratory of Surface Physics and Key
Laboratory of Micro and Nano Photonic Structures (Ministry of Education),
Fudan University, Shanghai, 200433, China
}
\author{\\Han Bao}%
\affiliation{%
 Department of Physics, State Key Laboratory of Surface Physics and Key
Laboratory of Micro and Nano Photonic Structures (Ministry of Education),
Fudan University, Shanghai, 200433, China
}%
\affiliation{QUANTUM, Johannes Gutenberg-Universit\"{a}t Mainz, 55128 Mainz, Germany}
\author{Heng Shen}%
 \affiliation{%
State Key Laboratory of Quantum Optics and Quantum Optics Devices,
Institute of Opto-electronics, Shanxi University, Taiyuan, Shanxi,
030006, China
}%
\affiliation{%
Collaborative Innovation Center of Extreme Optics, Shanxi University, Taiyuan 030006, China
}%
\author{Liantuan Xiao}%
\affiliation{%
State Key Laboratory of Quantum Optics and Quantum Optics Devices,
Institute of Laser Spectroscopy, Shanxi University, Taiyuan, Shanxi,
030006, China
}%
\affiliation{%
Collaborative Innovation Center of Extreme Optics, Shanxi University, Taiyuan 030006, China
}%
\author{Suotang Jia}%
\affiliation{%
State Key Laboratory of Quantum Optics and Quantum Optics Devices,
Institute of Laser Spectroscopy, Shanxi University, Taiyuan, Shanxi,
030006, China
}%
\affiliation{%
Collaborative Innovation Center of Extreme Optics, Shanxi University, Taiyuan 030006, China
}%
\author{ Mingfeng Wang}%
 \email{mfwang@wzu.edu.cn}
\affiliation{%
 Department of Physics, Wenzhou University, Zhejiang 325035, China
}%
\author{Yanhong Xiao}
\email{yxiao@fudan.edu.cn}
\affiliation{%
State Key Laboratory of Quantum Optics and Quantum Optics Devices,
Institute of Laser Spectroscopy, Shanxi University, Taiyuan, Shanxi,
030006, China
}%
\affiliation{%
Collaborative Innovation Center of Extreme Optics, Shanxi University, Taiyuan 030006, China
}%
\affiliation{%
Department of Physics, State Key Laboratory of Surface Physics and Key
Laboratory of Micro and Nano Photonic Structures (Ministry of Education),
Fudan University, Shanghai, 200433, China
}%
\maketitle
\setcounter{equation}{0}
\setcounter{figure}{0}
\renewcommand{\theequation}{S\arabic{equation}}
\renewcommand{\thefigure}{S\arabic{figure}}
\renewcommand{\thetable}{S\Roman{table}}
\renewcommand{\bibnumfmt}[1]{[S#1]}
\renewcommand{\citenumfont}[1]{S#1}

\section{1. Basic Theory}

\subsection{1.1. Atom-light interaction}

Our experiment relies on the interaction between a relatively strong optical pulse propagating in the $z$ direction and a hot $^{87}$Rb atomic sample.
The optical pulse probes the $D_2$ line, coupling the ground state $5\mathrm{S}_{1/2},F=2$ to the excited states in $5\mathrm{P}_{3/2}$.
In the case of far-off-resonant interaction, the excited levels can be adiabatically eliminated, which results in the effective Hamiltonian for the atom-light system \cite{BrainPHD}
\begin{eqnarray}\label{eq_Hint}
            \hat{H}_\mathrm{int}&=&-  \frac{\hbar c \Gamma}{8 A \Delta} \frac{\lambda^2}{2 \pi} \int_0^L\left\{a_0 \hat{\Phi}(z, t)+a_1 \hat{S}_z(z, t) \hat{j}_z(z, t)\right.
        \nonumber\\&&+a_2\left[\hat{\Phi}(z, t) \hat{j}_z^2(z, t)-\hat{S}_{-}(z, t) \hat{j}_{+}^2(z, t)\right.
         \left.\left.-\hat{S}_{+}(z, t) \hat{j}_{-}^2(z, t)\right]\right\} \rho A \text{d} z,
  \end{eqnarray}
where $\hat{\Phi}(z, t)$ represents photon flux, and $\hat{S}_{\pm} = \hat{S}_y \pm i\hat{S}_z$ are raising and lowering operators for light, which convert left-circularly-polarized $\sigma^-$ photons into right-circularly-polarized $\sigma^+$ photons or vice versa, and $\hat{j}_{\pm} = \hat{j}_y \pm i\hat{j}_z$ are the raising and lowering operators for the atoms.
$A = 7~\mathrm{mm}\times 7~\mathrm{mm}$ denotes the cross-section area of the laser beam and $L = 20~\mathrm{mm}$ represents the length of the cell in our experiment.
$\rho$ is the atomic density, $\hbar$ is the Plank constant, $c$ is the speed of light, and hereafter we take $\hbar = c = 1$.
$\lambda = 780~\mathrm{nm}$ and $\Gamma = 2\pi \times 6.07~\mathrm{MHz}$ are the wavelength of the probe light and decay rate of the excited state, respectively.
Here we neglect the structure of the laser beam profile and the Doppler broadening effect for simplicity.
The first scalar term of the Hamiltonian (\ref{eq_Hint}) causes DC stark shift for all ground state sub-levels, which can be neglected in the following calculation.
The second vector term causes a rotation of the Stokes vector $\mathbf{S}$ and the spin vector $\mathbf{J}$ around the $z$ axis, with which one may realize quantum non-demolition (QND) measurement of spin variables \cite{NP23, Nature.581.7807}.
The third term is the tensor part,  where the photon flux  $\hat{\Phi}(z, t)$ can be treated as a $c$-number, and the quadratic term $\hat j_z^2$ is induced by the second order Stark shift, leading to an effect known as one-axis twisting (OAT) evolution of the single spin, which enables squeezing of the internal spin \cite{PhysRevA.47.5138}.
The $\hat S_-\hat j_+^2$ term can flip a $\sigma^+$ photon into a $\sigma^-$ photon, and, at the same time, the atomic angular momentum along $z$ is raised by an amount of $2\hbar$.
The interaction coefficients $a_0$, $a_1$ and $a_2$ are given by \cite{BrainPHD}
\begin{equation}\label{eq_a1a2}
    \begin{aligned}
        a_0 &= \frac{\sqrt{2}}{20}\left(\frac{1}{1-\Delta_{13}/\Delta} +\frac{15}{1-\Delta_{23}/\Delta } +24\right), \\
        a_1 &= \frac{\sqrt{2}}{100}\left(-\frac{15}{1-\Delta_{13}/\Delta} -\frac{25}{1-\Delta_{23}/\Delta } +140\right), \\
        a_2 &= \frac{\sqrt{2}}{40}\left(\frac{1}{1-\Delta_{13}/\Delta} -\frac{5}{1-\Delta_{23}/\Delta} +4\right),
    \end{aligned}
\end{equation}
where $\Delta_{13}= 2\pi \times 423.60~\text{MHz}$ and $\Delta_{23}= 2\pi \times 266.65~\text{MHz}$ are the energy splitting between sublevels $|F'=1\rangle$ and $|F'=3\rangle$ and between the sublevels $|F'=2\rangle$ and $|F'=3\rangle$ of the excited state $5^2\mathrm{P}_{3/2}$, respectively.
$\Delta$ is the detuning of the probe light from the transition $5^{2}\mathrm{S}_{1/2},F=2 \rightarrow 5^{2}\mathrm{P}_{3/2},F'=3$. It is red detuning when $\Delta>0$ and blue detuning when $\Delta<0$.
Here we choose the red detuning to implement the spin squeezing, and the reason for this choice will be given in the following paragraphs.

\begin{figure}[htbp]
    \centering
    \includegraphics[width=0.75\linewidth]{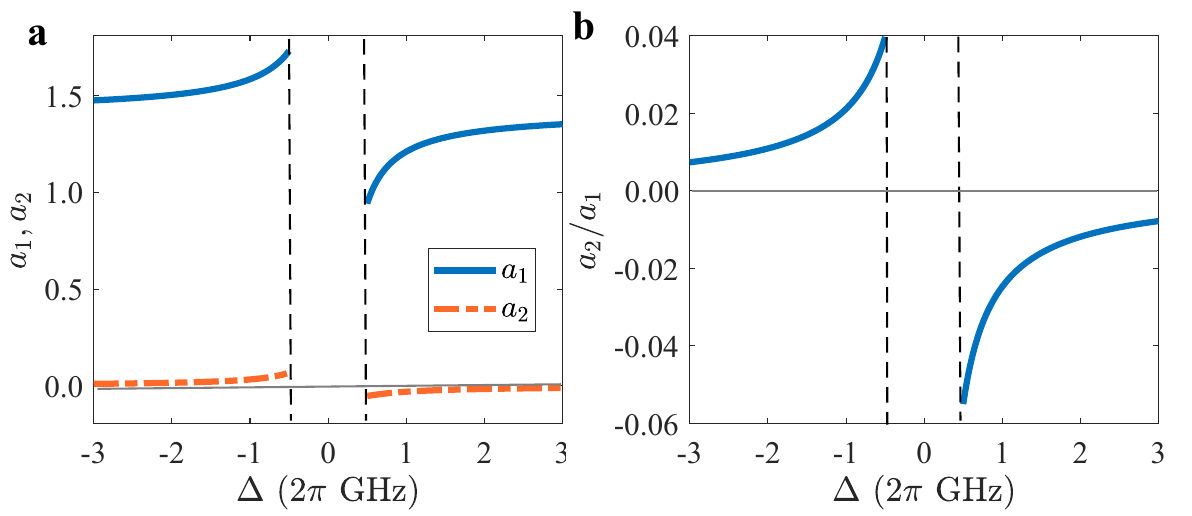}
    \caption{(a) The parameters $a_1$, $a_2$ and (b) the ratio $a_2/a_1$ vs. detuning $\Delta$, including red-detuning $\Delta>0$ and blue-detuning $\Delta<0$.
    The regions near resonance $-2\pi \times 500~\mathrm{MHz} < \Delta < 2\pi \times 500~\mathrm{MHz}$ are not depicted.}
    \label{fig_a1a2}
\end{figure}

In our experiment, the atomic sample is initially polarized along $x$ by pumping all the atoms to the Zeeman sub-level $|F=2, m_F=-2\rangle$ such that $\left\langle \hat j_x \right\rangle\approx -2$.
In this case, under the condition of weak atomic excitation, it is a good approximation to consider only the density matrix operators $\ket{-2}\bra{-2} (=\ket{F=2,m_F=-2}\bra{F=2,m_F=-2})$, $\ket{-1}\bra{-1}$, $\ket{-1}\bra{-2}$, and $\ket{-2}\bra{-1}$ \cite{BrainPHD}.
Then the spin components of a single atom can be approximated as ${\hat j _x} \approx  - 2\left| { - 2} \right\rangle \langle  - 2| - \left| { - 1} \right\rangle \langle  - 1|$, ${\hat j _y} \approx \left| { - 2} \right\rangle \langle  - 1| + \left| { - 1} \right\rangle \langle  - 2|$, ${\hat j_z} \approx i(\left| { - 2} \right\rangle \langle  - 1| - \left| { - 1} \right\rangle \langle  - 2|)$, which leads to $\hat j_y\hat j_x+\hat j_x\hat j_y\approx-3\hat j_y$, and thus we arrive at
\begin{eqnarray}
{{\hat H}_{{\rm{int}}}} &\simeq&  - \frac{{\Gamma }}{{8A\Delta }}\frac{{{\lambda ^2}}}{{2\pi }}\int_0^L {\left\{ {{a_0}\hat \Phi (z,t) + {a_1}{{\hat S}_z}(z,t){{\hat j}_z}(z,t)} \right.} \nonumber\\
 &&+ {a_2}\left[ {\hat \Phi (z,t)\hat j_z^2(z,t) + 2{{\hat S}_x}(z,t)\hat j_y^2} \right.(z,t)\left. {\left. { - 8{{\hat S}_x}(z,t) +6{{\hat S}_y}(z,t){{\hat j}_y}(z,t)} \right]} \right\}\rho A{\rm{d}}z.\label{eqs3}
\end{eqnarray}
By appropriate choice of the polarization of the probe light, the atomic quadratic interactions $\hat j_{z}^2,\hat j_{y}^2$ in the second line of Eq.~(\ref{eqs3}) can establish entanglement between the nuclear spin and the electronic spin, and thus produce spin squeezing \cite{PhysRevLett.101.073601}.
The squeezing direction of such interactions, however, typically varies with time, which might decrease the efficiency of squeezing creation of our proposed mechanism presented in the main text.
To cancel out the influence of these quadratic terms, we assume that the probe light is linearly polarized along $x$ axis ($\pi$ polarization along $x$ axis) such that $\hat{S}_x = \hat\Phi/2$, which results in $\hat \Phi \hat j_z^2 + 2{{\hat S}_x}\hat j_y^2 \simeq \hat \Phi ( {\hat j_z^2 + \hat j_y^2} ) = 6\hat \Phi  - \hat \Phi \hat j_x^2$.
Since the photon flux $\hat\Phi$ is usually treated as a $c$-number \cite{BrainPHD}, the atomic quadratic term $\hat j_x^2$ only shifts the Larmor frequency, while the light term $\hat S_x$ causes a rotation of the light state in the $\hat S_y$-$\hat S_z$ plane and is negligible.
As a result, we obtain
\begin{eqnarray}
{{\hat H}_{{\rm{int}}}} \simeq  \frac{\kappa}{\sqrt{\Phi N_A}} \int_0^L {\left[ {{{\hat S}_z}(z,t){{\hat j}_z}(z,t) -\zeta^2{{\hat S}_y}(z,t){{\hat j}_y}(z,t)} \right]}\rho A{\rm{d}}z,\label{seq4}
\end{eqnarray}
where the coupling strength $\kappa = -\Gamma\lambda^2 a_1 \sqrt{\Phi N_A}/(16\pi A\Delta)$ and $\zeta^2=-6a_2/a_1 $.
Next, we utilize the Holstein-Primakoff approximation \cite{PhysRev.58.1098} to map atomic and light operators to Bosonic creation and annihilation operators, by defining \cite{Wasilewski:09}
\begin{eqnarray}
{{\hat x}_L} &=& \frac{{{{\hat S}_y}}}{{\sqrt{\Phi/2}}} = \frac{1}{{\sqrt 2 }}\left(\hat a + {{\hat a}^\dag }\right),{{\hat p}_L} = \frac{{{{\hat S}_z}}}{{\sqrt{\Phi/2}}} = \frac{{ - i}}{{\sqrt 2 }}\left(\hat a - {{\hat a}^\dag }\right).\label{seq5}\\
{{\tilde {\hat x}}_A} &=&  - \frac{{{{\tilde {\hat j}}_y}}}{{\sqrt {2{N_A}/L} }} = \frac{1}{{\sqrt 2 }}\left(\tilde {\hat b} + {{\tilde {\hat b}^\dag} }\right),{{\tilde {\hat p}}_A} = \frac{{{{\tilde {\hat j}}_z}}}{{\sqrt {2{N_A}/L} }} = \frac{{ - i}}{{\sqrt 2 }}\left(\tilde {\hat b} - {{\tilde {\hat b}^\dag} }\right).\label{seq6}
\end{eqnarray}
The atomic annihilation operator $\tilde{\hat b} = \sum_{i = 1}^{{N_A}/L} \left| { - 1} \right\rangle_i \langle  - 2| /\sqrt{{N_A}/L}$ can annihilate an atomic excitation within an atomic slice around a given $z$ and at a certain time $t$, satisfying $[\tilde{\hat b}(z,t),\tilde{\hat b}^\dag(z',t)]=\delta(z-z')$, while the light annihilation operator $\hat a$ annihilates a photonic excitation around a given $z$ satisfying $[\hat a(z,t),\hat a^\dag(z,t')]=\delta(t-t')$.
With Eqs. (\ref{seq5}) and (\ref{seq6}), the Hamiltonian (\ref{seq4}) can be re-expressed as
\begin{eqnarray}
{{\hat H}_{{\mathop{\rm int}} }} &=& \frac{\kappa}{\sqrt{L}} \int_0^L {\left[ {{{\tilde {\hat p}}_A}(z,t){{\hat p}_L}(z,t) + {\zeta ^2}{{\tilde {\hat x}}_A}(z,t){{\hat x}_L}(z,t)} \right]\mathrm{d}z}
 \nonumber\\
&=& \int_0^L {\left[ \tilde{\hat b}^\dagger \left( \mu_+ \hat{a} - \mu_- \hat{a}^\dagger \right) + \mathrm{h.c.} \right]\mathrm{d}z} ,\label{eqs7}
\end{eqnarray}
where ${\mu _ \pm } =  \kappa (1 \pm {\zeta ^2})/2\sqrt{L}$.
This Hamiltonian describes a beam splitter (BS) type interaction between the atomic mode $\tilde{\hat b}$ and the  photonic Bogoliubov mode $\hat{\Gamma} \propto \mu_+ \hat{a} - \mu_- \hat{a}^\dagger$.
The Hamiltonian (\ref{eqs7}) serves as the starting point for our current investigation.
Note that, when $\mu_+ = \mu_-$, the interaction Hamiltonian is exactly the well-known QND interaction, while our current scheme relies on an unbalanced interaction, that is, $\mu_+ > \mu_-$, as analyzed in detail in the main text.
To engineer such an unbalanced interaction, it is required that $\zeta ^2>0$, which can be ensured by choosing the detuning $\Delta>0$ (red-detuning), as shown in Fig.~\ref{fig_a1a2}(b).

\subsection{1.2. The stroboscopic interaction}
In reality, a homogenous magnetic field is usually applied along the $x$ direction to hold the macroscopic spin to reduce the dephasing, which changes the interaction (\ref{eqs7}) into
\begin{equation}
{\hat H} \propto \Omega {\hat b^\dag }\hat b - {\mu _ - }{\hat b^\dag }{\hat a^\dag } + {\mu _ + }{\hat b^\dag }\hat a + {\rm{h}}.{\rm{c}},\label{s8}
\end{equation}
where $\Omega$ is the Larmor frequency and we have discarded the tilde on $\hat{b}$ for simplicity. The interaction term (the second and third terms) alone can deterministically produce spin squeezing along the quadrature direction $\hat p_A$. The first term of Eq. (\ref{s8}) alone leads to the equation of motion, $d\hat p_A /dt=-\Omega \hat x_A$, indicating that information about $\hat x _A$ will be brought back to $\hat p_A$ by the $\hat b^{\dag}\hat b$ interaction. Since the interaction term of Eq. (\ref{s8}) leads to the anti-squeezing of $\hat x_A$, its noise is large and therefore the variance of $\hat p_A$ is increased, destroying the spin squeezing, which we call the ``back-action'' noise.

\begin{figure}[htbp]
\centering
\includegraphics[width=0.98\linewidth]{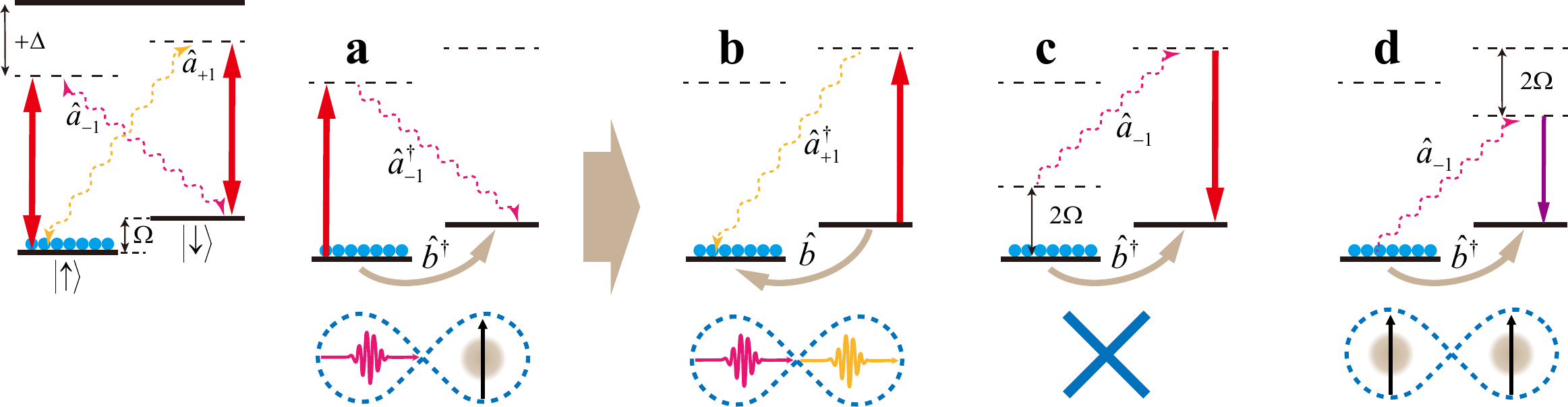}
\caption{Left: Level diagram of the double-Raman type interaction without stroboscopic, $\hat{a}_{-1}$ and $\hat{a}_{+1}$ describe the Stokes and anti-Stokes scattering processes, respectively.
Right: Step-by-step squeezing generation process. (a): The entangled Stokes-photon-spin pair is produced by the TMS process $\hat{b}^\dagger \hat{a}_{-1}^\dagger$. (b): The photon-spin entanglement is transferred to the photon-photon entanglement through the BS interaction $\hat{b} \hat{a}_{+1}^\dagger$. (c): It is difficult to transfer photon-spin entanglement onto the spin due to the existence of two-photon detuning $2\Omega$. (d): The spin-spin entanglement can be established via the BS interaction $\hat{b}^\dagger \hat{a}_{-1}$ with the assistance of the lower sideband light (with frequency $\omega_0-2\Omega$) created by the stroboscopic driving. }
\label{figs_moti}
\end{figure}

To find the way to to reduce the influence of such unwanted back-action noise, we write the Hamiltonian (\ref{s8}) in the frequency domain
\begin{equation}
\hat{H}_\mathrm{int} \propto -\mu_- \hat{b}^\dagger \hat{a}_{-1}^\dagger  +\mu_+ \hat{b}^\dagger \hat{a}_{+1} + \mathrm{h.c.},\label{s9}
\end{equation}
where $\hat{a}_{k}$ describes the Stokes and anti-Stokes light with frequency $\omega_0 + k\Omega$, as shown in the main text.
This kind of interaction can be understood as a double-Raman process as depicted in the left of Fig.~\ref{figs_moti}. The Stokes photons couple to the atoms via a two-mode-squeezing (TMS) interaction $\hat{b}^\dagger \hat{a}_{-1}^\dagger + \mathrm{h.c.}$, while the anti-Stokes photons interact with atoms via a beam-splitter (BS) process $\hat{b}^\dagger \hat{a}_{+1} + \mathrm{h.c.}$.
The atom-light interaction (\ref{s9}) has a clear physical meaning: the TMS interaction $\hat{b}^\dagger \hat{a}_{-1}^\dagger$ creates a Stokes photon in the lower sideband $\omega_0-\Omega$ in a forward direction, and, at the same time, a spin is flipped;
clearly, an entangled photon-spin pair is created, as shown in Fig.~\ref{figs_moti}(a).
Once the photon-spin entanglement appears, the spin excitation can be annihilated via the BS interaction $\hat{b} \hat{a}_{+1}^\dagger$ and, at the same time, an anti-Stokes photon is generated, as depicted in Fig.~\ref{figs_moti}(b). As a result, the quantum process (a)$\rightarrow$(b) transfers the photon-spin entanglement onto the photon-photon entanglement, leading to two-mode light squeezing. However, the quantum process (a)$\rightarrow$(c) cannot induce spin squeezing, as the generated photon-spin entanglement cannot be swapped onto the spins due to the presence of two-photon detuning ($=2\Omega$) between the emitted Stokes photons and the strong driving field, as shown in Fig.~\ref{figs_moti}(c).

To address this problem, we proposed to use a stroboscopic driving with frequency $2\Omega$, leading to multiple sidebands with frequency separation $2\Omega$, as shown in Fig. 1(b) in the main text.
One can see immediately that the lower sideband light with frequency $\omega_0-2\Omega$ can stimulate the emission of anti-Stokes photon with frequency $\omega_0-\Omega$, which is the same as the Stokes photon stimulated by the carrier light.
This implies that as soon as the Stokes photon $\hat{a}_{-1}$ is emitted, it has high probability to be absorbed by the atoms via the BS interaction $\hat{b}^\dagger \hat{a}_{-1}$, which swaps the excitation of $\hat{a}_{-1}$ onto the atomic mode $\hat b$ [see Fig. ~\ref{figs_moti}(d)], creating spin-spin entanglement and thus the spin squeezed state, corresponding to the quantum process (a)$\rightarrow$(d).
On the other hand, its Hermitian conjugate $\hat{b} \hat{a}_{-1}^\dagger$ enables the excitation transfer from the collective atomic mode to the light mode $\hat{a}_{-1}$, generating single-mode light squeezed state.
Such a qualitative analysis can be extended to the case of higher sidebands, predicting that the light modes $\hat{a}_{k}$ with $k\in \rm{odd}$ will also exhibit squeezing properties, which forms a quantum system that is in close analogy to the squeezed vacuum reservoir \cite{scully_quantum_1997} (see the main text for more details).
Therefore, with the assistance of the stroboscopic driving, it is possible to realize concurrent light and spin squeezing.

\subsection{1.3. Propagation equations}
In our experiment a homogeneous magnetic field is applied along the $x$-direction, which gives an additional term $\hat H_A=\Omega \tilde{\hat b}^\dag\tilde{\hat b}$ to the Hamiltonian.
Furthermore, our approach relies on a stroboscopic dynamic, which involves breaking up the probe pulse into a train of short optical pulses, leading to a time dependent coupling strength $\kappa(t)=\kappa\phi(t)$, where the temporal-profile function $\phi(t)$ is assumed to keep the peak amplitude of $\kappa(t)$ unchanged.
We therefore obtain the total Hamiltonian $\hat H_\mathrm{tot}(t)=\hat H_A+\phi(t)\hat H_{\rm{int}}$.
Using this Hamiltonian, to derive equations of motion we use the Heisenberg equations $\partial_t\hat \vartheta_A=-i[\hat\vartheta_A,\hat H_T]$ ($\vartheta\in\{x,p\}$) and the Maxwell-Bloch equations $(\partial_t+\partial_z)\hat \vartheta_L=-i[\hat\vartheta_L,\hat H_\mathrm{tot}]$ for atomic and light operators, respectively, yielding
\begin{subequations}
\label{eq:whole1}
\begin{eqnarray}
\left( {\frac{\partial }{{\partial t}} + \frac{\partial }{{\partial z}}} \right){{\hat x}_L}(z,t) &=& \frac{\kappa}{\sqrt{L}} \phi \left( t \right){{\tilde {\hat p}}_A}(z,t),\label{s8a}\\
\left( {\frac{\partial }{{\partial t}} + \frac{\partial }{{\partial z}}} \right){{\hat p}_L}(z,t) &=& -\frac{{\zeta ^2}\kappa}{\sqrt{L}} \phi \left( t \right){{\tilde {\hat x}}_A}(z,t),\label{s8b}\\
\frac{{\partial {{\tilde {\hat x}}_A}(z,t)}}{{\partial t}} &=& \Omega {{\tilde {\hat p}}_A}(z,t) + \frac{\kappa}{\sqrt{L}} \phi \left( t \right){{\hat p}_L}(z,t),\label{s8c}\\
\frac{{\partial {{\tilde {\hat p}}_A}(z,t)}}{{\partial t}} &=&  - \Omega {{\tilde {\hat x}}_A}(z,t) - \frac{{\zeta ^2}\kappa}{\sqrt{L}} \phi \left( t \right){{\hat x}_L}(z,t).\label{s8d}
\end{eqnarray}
\end{subequations}
Next, we neglect retardation effects of light \cite{BrainPHD}, that is, the dynamics
on the time scale $L/c$ of the light propagation through the cell will not be calculated.
The $t$-differentiation $\partial_t$ in Eqs. (\ref{s8a}) and (\ref{s8b}) can then be left out.
After integrating Eqs. (\ref{s8a}) and (\ref{s8b}) on both sides formally with respect to $z$, we have
\begin{subequations}
\label{eq:whole2}
\begin{eqnarray}
{{\hat x}_L}(z,t) &=& {{\hat x}_L}(0,t) + \frac{\kappa}{\sqrt{L}} \phi \left( t \right)\int_0^z {{{\tilde {\hat p}}_A}(\tilde z,t)\mathrm{d}\tilde z} ,\label{s9a}\\
{{\hat p}_L}(z,t) &=& {{\hat p}_L}(0,t) - \frac{{\zeta ^2}\kappa}{\sqrt{L}} \phi \left( t \right)\int_0^z {{{\tilde {\hat x}}_A}(\tilde z,t)\mathrm{d}\tilde z} ,\label{s9b}
\end{eqnarray}
\end{subequations}
Substituting Eqs.~(\ref{s9a}) and (\ref{s9b}) into Eqs.~(\ref{s8c}) and (\ref{s8d}) we find
\begin{subequations}
\label{eq:whole3}
\begin{eqnarray}
\frac{{\partial {{\hat x}_A}(t)}}{{\partial t}} &=& \Omega {{\hat p}_A}(t) + \kappa \phi \left( t \right){{\hat p}_L}(0,t) - \frac{{\zeta ^2}{\kappa ^2}}{\sqrt{L}}{\phi ^2}\left( t \right)\int_0^L {\left( {1 - z/L} \right){{\tilde {\hat x}}_A}(z,t)\mathrm{d}z} ,\\
\frac{{\partial {{\hat p}_A}(t)}}{{\partial t}} &=&  - \Omega {{\hat x}_A}(t) + {\zeta ^2}\kappa \phi \left( t \right){{\hat x}_L}(0,t) - \frac{{\zeta ^2}{\kappa ^2}}{\sqrt{L}}{\phi ^2}\left( t \right)\int_0^L {\left( {1 - z/L} \right){{\tilde {\hat p}}_A}(\tilde z,t)\mathrm{d}z},
\end{eqnarray}
\end{subequations}
where $\hat \vartheta_A=\int_{0}^{L}\tilde{\hat \vartheta}_A(z)\mathrm{d}z/\sqrt{L}$ denotes the collective atomic mode.
The last terms of Eq. (\ref{eq:whole3}) indicate that information about atoms themselves is brought back by light, inducing a nonlinear spin-spin evolution, which is the origin of spin squeezing.
The spatially varying functions $1-z/L$ reveals that the atomic information brought back by light from the front of the sample is significantly larger compared to the rear of the sample.
Such a result can be qualitatively understood as follows:
considering a particular piece of light in the probe pulse, once it enters the cell, it would first pick up the atomic information at $z=0$;
as it propagates along the cell (its propagation distance is $L$), information about the first atomic spatial mode $\tilde{\hat b}(0)$ will gradually be mapped onto the incoming spatial modes $\tilde{\hat b}(z)(0<z\leq L)$ via $\hat H_{\rm{int}}$ interaction;
therefore, the closer the atomic spatial modes to $z=0$, the higher the chance it has to be mapped onto the subsequent atomic spatial modes.
Consequently, it seems that there is no chance that the last spatial mode $\tilde{\hat b}(L)$ gets recorded by the atoms since $1-L/L=0$, but here it is not the case.
Due to the thermal motion, each atom can move in and out different spatial modes many times (e.g., our experiment condition leads to about $40$ wall collisions) during the probe process.
As a result, different pieces of light see different atomic spatial modes, and, after averaging over the whole pulse, one approximately has $\int_0^L {( {1 - z/L} ){{\tilde {\hat x}}_A}(z,t)\mathrm{d}z}  \simeq {{\hat x}_A}(t)\sqrt{L}/2$ \cite{Wasilewski:09}, which results in
\begin{subequations}
\label{eq:whole4}
\begin{eqnarray}
\hat x_L^{{\rm{out}}}(t) &=& \hat x_L^{{\rm{in}}}(t) + \kappa \phi (t)\left[ { - {{\hat x}_A}(t)\sin (\Omega t) + {{\hat p}_A}(t)\cos (\Omega t)} \right],\label{s11a}\\
\hat p_L^{{\rm{out}}}(t) &=& \hat p_L^{{\rm{in}}}(t) - {\zeta ^2}\kappa \phi (t)\left[ {{{\hat x}_A}(t)\cos (\Omega t) + {{\hat p}_A}(t)\sin (\Omega t)} \right],\label{s11b}\\
\partial_t{{\hat x}_A}(t) &=&  - \zeta^2\kappa^2 \phi^2 (t){{\hat x}_A}(t) /2 + \kappa \phi (t)\hat p_L^{{\rm{in}}}(t)\cos (\Omega t) + {\zeta ^2}\kappa \phi (t)\hat x_L^{{\rm{in}}}(t)\sin (\Omega t),\label{s11c}\\
\partial_t{{\hat p}_A}(t) &=&  - \zeta^2\kappa^2\phi^2 (t){{\hat p}_A}(t) /2 + \kappa \phi (t)\hat p_L^{{\rm{in}}}(t)\sin (\Omega t) - {\zeta ^2}\kappa \phi (t)\hat x_L^{{\rm{in}}}(t)\cos (\Omega t),\label{s11d}
\end{eqnarray}
\end{subequations}
where we have changed to a rotating frame $\hat b\rightarrow e^{i\Omega t}\hat b$.
For the light operators, the newly defined operators $\hat \vartheta^{\rm{in}}(t)=\hat \vartheta(0,t)$ and $\hat \vartheta^{\rm{out}}(t)=\hat \vartheta(L,t)$ represent the light spatial modes before and after the interaction with the sample, respectively.
Next, this set of equations will be solved to see concurrent light and spin squeezing.

\subsection{1.4. Theoretical results for concurrent light and spin squeezing}

In the main text, we explained that both light and spin are squeezed in the $\hat p$ quadrature, i.e., along the $S_z$ and $J_z$ direction respectively. Here we provide another physical picture. In terms of the atomic spin $J$ and light Stokes operator $S$, the Hamiltonian can be also written as $\hat{H} = \hat{J} _z \hat{S} _z - A \hat{J} _y \hat{S} _y$ [Eq.~(\ref{eqs7})]. The first term $\hat{J} _z \hat{S} _z$ is an effective Faraday rotation for both spin and light, and A is a positive coefficient less than one. For light, after $\hat{J} _z \hat{S} _z$ interaction, $\hat{S} _y$ will carry the information of $\hat{J} _z$.
Substituting $\hat{S} _y$ in the second term by $\hat{J} _z$, one has a term proportional to $\hat{J} _y \hat{J} _z$. This $\hat{J} _y \hat{J} _z$ term will squeeze the spin along $z$ direction with negative $J_x$. Because of the symmetry between $J$ and $S$, or applying the same reason for $\hat{J} _y$, one can see that light is squeezed along $\hat{S} _z$.

Next, we derive the dependence of squeezing on experiment parameters.

Since the temporal profile function $\phi(t)$ in Eq.~(\ref{eq:whole4}) changes periodically over time [see Fig.~1(b)], it is convenient to apply the Fourier expansion \cite{PhysRevLett.106.143601}
\begin{eqnarray}\label{stro}
   \phi(t)=\sum_{n=-\infty}^{\infty}\mathcal{A}_n e^{i2n\Omega t}, \text{with}~ \mathcal{A}_n = d~\mathrm{sinc}(\pi n d),\label{s12}
\end{eqnarray}
where $2\Omega$ is the stroboscopic frequency, $d$ is the duty cycle, and $\mathrm{sinc}(x) = \sin(x)/x$ is the sinc function.
Obviously, the optical spectral distribution of the strong field is a series of peaks located at $\omega=\pm 2n\Omega$, with an envelop determined by the sinc function, and peak separation $2\Omega$ [see Fig.~1(b) in the main text], which is also known as the optical frequency comb. Then, the atom-light interaction process can also be understood as an atomic ensemble interacting simultaneously with a group of $x$-polarized classical driving pulses which are in a single temporal and spatial mode and whose carrier frequencies lie at  $\omega_{\pm n}=\omega_0\pm 2n\Omega$ (with $\omega_0$ being the carrier frequency of the non-stroboscopic light). As is shown in the inset of Fig.~1(a) in the main text, because of the Zeeman splitting,
these probe sideband light mode will stimulate the emission of Stokes or anti-Stokes photons that have different frequency.
The probe light with frequency $\omega_0 + 2n\Omega$ couples to the anti-Stokes photon with frequency $\omega_0 + (2n-1)\Omega$ by $\tilde{\hat{b}}^\dagger \hat{a}$, and Stokes photon with frequency $\omega_0 + (2n+1)\Omega$ by $\tilde{\hat{b}}^\dagger \hat{a}^\dagger$.
In this picture, if we define the photon creation operator as $\hat{a}_{k}$ with frequency $\omega_0 + k\Omega$, then we can write down the Hamiltonian as:
\begin{eqnarray}\label{eq1}
\hat{H}(t) &\propto& \sum_{n=-\infty}^\infty \mathcal{A}_n \big( \mu_+ \tilde{\hat{b}}^\dagger {\hat{a}_{2n+1}} - \mu_- \tilde{\hat{b}}^\dag \hat{a}_{2n-1}^\dag + \mathrm{h.c.}\big) \nonumber\\
&=& \tilde{\hat{b}}^\dag \sum_{k=-\infty}^\infty \big( \mathcal{A}_{(k-1)/2}  \mu_+ \hat{a}_{k} - \mathcal{A}_{(k+1)/2} \mu_- \hat{a}_{k}^\dag\big) + \mathrm{h.c.} \nonumber\\
&=& {{\hat b}^\dag }\sum\limits_{k =  - \infty }^\infty  {\mathcal{R}_k\left( {{{\hat a}_{k}}\cosh r_k - \hat a_{k}^\dag \sinh r_k} \right)}  + {\rm{h}}{\rm{.c}}{\rm{.}},
\end{eqnarray}
where $k$ is odd, ${\mathcal{R}_k} = {(\mathcal{A}_{(k - 1)/2}^2\mu _ + ^2 -\mathcal{ A}_{(k + 1)/2}^2\mu _ - ^2)^{1/2}}$, ${r_k} = \ln [({\mathcal{A}_{(k - 1)/2}}{\mu _ + } + {\mathcal{A}_{(k + 1)/2}}{\mu _ - })/({\mathcal{A}_{(k - 1)/2}}{\mu _ + } - {\mathcal{A}_{(k + 1)/2}}{\mu _ - })]$, and  Eq.~(\ref{eq1}) here is Eq.~(1) in the main text.

From Eq.~(\ref{s12}), one may directly calculate $\phi^2(t)=\sum_{n=0}^\infty\mathcal{A}_n^2+\rm{oscillating~terms}$.
Since we are interested in the dynamic that is time-averaged over a period much longer than the period of Larmor precession (i.e. the pulse time duration $T\gg 2\pi/\Omega$), the oscillating terms can be neglected.
As a result, in the evolution equations (\ref{s11c}) and (\ref{s11d}), the prefactor of the first term can be approximated as $\gamma_s \equiv \zeta^2\kappa^2 \phi^2 (t) /2 \approx d \zeta^2\kappa^2 /2$.
Then, integrating on both sides of equations (\ref{s11c}) and (\ref{s11d}) yields
\begin{eqnarray}
\hat{x}_A(t) &=& \hat{x}_A(0)e^{-\gamma_s t} + \kappa\int_{0}^{t}\mathrm{d}\tau~ e^{-\gamma_s(t-\tau)}\phi(\tau) \hat{p}_L^\mathrm{in}(\tau)\cos(\Omega \tau) + \zeta^2\kappa\int_{0}^{t}\mathrm{d}\tau~ e^{-\gamma_s(t-\tau)}\phi(\tau)\hat{x}_L^\mathrm{in}(\tau)\sin(\Omega \tau), \label{s13}\\
\hat{p}_A(t) &=&\hat{p}_A(0)e^{-\gamma_s t}+\kappa\int_{0}^{t}\mathrm{d}\tau~ e^{-\gamma_s(t-\tau)}\phi(\tau) \hat{p}_L^\mathrm{in}(\tau)\sin(\Omega \tau) - \zeta^2\kappa\int_{0}^{t}\mathrm{d}\tau~ e^{-\gamma_s(t-\tau)}\phi(\tau)\hat{x}_L^\mathrm{in}(\tau)\cos(\Omega \tau) .\label{s14}
\end{eqnarray}
We can see that $\gamma_s$ characterizes the decay of the initial state and its corresponding noise, so we call it squeezing rate.
Considering the fact that $\left\langle \hat{p}_\mathrm{A}(t) \right\rangle = 0$ and using the expectation values $\langle \hat{\vartheta}_L^{in}(t) \hat{\vartheta}_L^{in}(t') \rangle = \delta(t-t')/2$ of the input light field, the variance of $\hat p_A$ at time $T$ under the condition $\omega_m \gg \gamma_s$ can be derived:
\begin{eqnarray} \label{eqVt}
\mathrm{Var}( \hat{p}_A)_\mathrm{SSS} &=& \frac{1}{2} \left[e^{-2\gamma_s T} + \left(1-e^{-2\gamma_s T}\right) \mathcal{D}_{A} \right],\\
\mathcal{D}_{A} &=&  \frac{1- \mathrm{sinc}(\pi d)}{2} \zeta^{-2} + \frac{1+ \mathrm{sinc}(\pi d)}{2} \zeta^{2} ,\nonumber
\end{eqnarray}
with $\mathrm{Var}(x) = \left\langle x^2 \right\rangle - \left\langle x \right\rangle^2$.
One may derive the squeezing parameter $\xi_{A}^2 $ $= \mathrm{Var}(\hat{p}_{A})_\mathrm{SSS}/$ $\mathrm{Var}(\hat{p}_{A})_\mathrm{PNL}$ $=2\mathrm{Var}(\hat{p}_{A})_\mathrm{SSS}$, where  $\mathrm{Var}(\hat{p}_{A})_\mathrm{PNL} = 1/2$ in theory is the projection noise limit (PNL).
With a fixed $d$, it can also be expressed in the form $\xi_A^2(T) = b_1+b_2~e^{-2\gamma_s T}$ where $b_{1,2}$ are coefficients derived from Eq.~(\ref{eqVt}), which is an exponentially decreasing noise whose long time limit ($T\gg \gamma_s^{-1}$) goes to $\xi_{A}^2 \to \mathcal{D}_{A}$.
Obviously, this limit cannot be reached in realistic systems due to the finite value of macroscopic-spin decay time $T_1$.
Considering the impact of the decay of the macroscopic spin, the spin squeezing parameter of Wineland criterion \cite{PhysRevA.50.67} becomes
\begin{eqnarray} \label{eqVt1}
\xi_{A,W}^2(T) = e^{2T/T_1} \left[e^{-2\gamma_s T} +  \varepsilon\left(1-e^{-2\gamma_s T}\right) \mathcal{D}_{A} \right],
\end{eqnarray}
which is the result of Eq.~(2) in the main text, and can be simplified to $\xi_{A,W}^2(T) = e^{2T/T_1}\left(b_1+b_2~e^{-2\gamma_s T}\right)$, the fitting function of Fig.~3(a) in the main text.
We can also take squeezing time $T$ as a fixed parameter, then the squeezing parameter becomes $\xi_{A,W}^2(d) = c_1 + c_2~ \mathrm{sinc}(\pi d)$ where $c_{1,2}$ are coefficients derived from Eq.~(\ref{eqVt1}), which is the fitting function of Fig.~4(a) in the main text.

For spin variance along an arbitrary direction $\alpha$, we define the spin quadrature $\hat{q}_\alpha = \hat{p}_A \cos (\alpha/2) + \hat{x}_A \sin (\alpha/2)$.
Under a similar process, we can get the spin squeezing parameter the same as Eq.~(\ref{eqVt1}), with
\begin{equation*}
\mathcal{D}_{A} \rightarrow \mathcal{D}_{A}(\alpha) = \frac{1- \cos \alpha~\mathrm{sinc}(\pi d)}{2} \zeta^{-2} + \frac{1+ \cos \alpha~\mathrm{sinc}(\pi d)}{2} \zeta^{2}.
\end{equation*}
Taking squeezing time $T$ and duty cycle $d$ as fixed parameters, the squeezing parameter can be simplified to $\xi_{A,W}^2(\alpha) = d_1+d_2~\cos \alpha$, where $d_{1,2}$ are coefficients derived from Eq.~(\ref{eqVt1}), and is the fitting function of Fig.~3(b) in the main text.

\begin{figure}[htbp]
\centering
\includegraphics[width=0.5\linewidth]{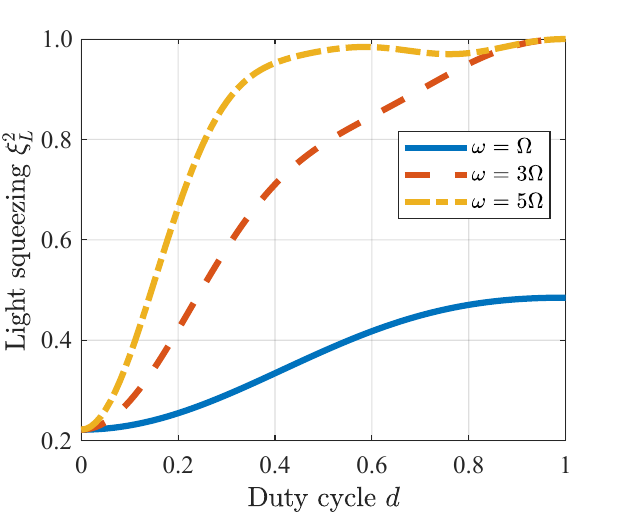}
\caption{Light squeezing at different frequency vs. duty cycle $d$.
Solid blue curve is the light squeezing at $\omega=\Omega$ ($k=1$), red dashed curve is at $\omega=3\Omega$  ($k=3$) and orange dash-dotted curve is at $\omega=5\Omega$  ($k=5$). We here take $\gamma_s T = 1, \zeta^2 = 0.1$, and the light is squeezed if $\xi_L^2 <1$.}
\label{figs2}
\end{figure}
In order to calculate light squeezing, we define the light spectrum $S(\omega) = \int_0^T\mathrm{d}t \int_{-t}^{T-t}\mathrm{d}\tau~ e^{i\omega\tau} \phi(t+\tau) \phi(t)$ $ \left\langle \hat{p}_{L}^\mathrm{out}(t+\tau)\hat{p}_{L}^\mathrm{out}(t) \right\rangle/ dT$, where $\omega$ is the detection frequency.
Then, using Eqs.~(\ref{s11a}) and (\ref{s11b}) one may derive the light squeezing spectrum of $\hat{p}_{L}^\mathrm{out}$ at different frequencies:
\begin{eqnarray}\label{eqSw}
        S(\omega)_\mathrm{LSS} &=& \frac{1}{2} \left\{1 - \sum_{k=-\infty}^{\infty} \frac{{\left[ \mathcal{D}_{L,\mathrm{corr}}(\frac{k-1}{2}) - \mathcal{D}_{L,\mathrm{QBA}}(\frac{k-1}{2}) \right] }}{{1 + {{\left[ {\omega  - k\Omega } \right]}^2/{\gamma_s ^2}}}}\right\},\label{s16}\\
        \mathcal{D}_{L,\mathrm{corr}}(n) &=& f_2 \alpha(n),\nonumber\\
        \mathcal{D}_{L,\mathrm{QBA}}(n) &=& \zeta^2 f_1 \alpha(n) + \left( f_2-f_1 \right) \bigg[  \frac{1}{2} \left( 1+\zeta^4 \right) \alpha(n) + \left( 1-\zeta^4 \right) \beta(n) \bigg],\nonumber\\
        \alpha(n) &=& \mathrm{sinc}^2(\pi n d) + \mathrm{sinc}^2[\pi (n+1) d],\nonumber\\
        \beta(n) &=& \mathrm{sinc}(\pi d)~\mathrm{sinc}(\pi n d)~\mathrm{sinc}[\pi (n+1) d],\nonumber
   \end{eqnarray}
where $k$ is odd integer and $f_1 = \left( 1-e^{-\gamma_s T} \right)^2/\gamma_s T, f_2 = 2-2\left( 1-e^{-\gamma_s T} \right)/\gamma_s T$.
The light squeezing parameter can be obtained by calculating the ratio of the noise of the light squeezed state (LSS) to the shot noise (SN), denoted as $\xi_L^2 (\omega)= S_\mathrm{LSS}(\omega)/S_\mathrm{SN}(\omega)$, where $S_\mathrm{SN}(\omega) = 1/2$ in theory and is the Eq.~(3) in the main text.
Taking squeezing time as a fixed parameter, one may acquire the light squeezing at different frequencies (different $k$) as $\xi_L^2(k,d) = 1 - e_1 \alpha[(k-1)/2,d] - e_2 \beta[(k-1)/2,d]$ where $e_{1,2}$ are coefficients derived from Eq.~(\ref{s16}), which is the fitting function of Fig.~2(b) (with fixed $d$) and Fig.~4(b) (with $k=1$ and then the fitting function can be simplified to $\xi_L^2(d) = h_1 + h_2 \mathrm{sinc}^2(\pi d)$) in the main text.
In Fig.~(\ref{figs2}), we plot $\xi_L^2(k,d)$ as a function of duty cycle $d$, indicating that
(i) the amount of squeezing decreases with the orders $k$ of the sidebands, and
(ii) smaller duty cycle can benefit light squeezing, and
(iii) for the first-order sideband ($k=1, \omega=\Omega$), light squeezing still exists even without stroboscopic ($d=1$).
These theoretical predictions are consistent with experimental findings presented in the main text.
With fixed $d$ and $k$, one may also express the light squeezing at different squeezing time $T$ as $\xi_L^2(T) = 1 - g_1 f_1(T) - g_2 f_2(T)$ where $g_{1,2}$ are coefficients derived from Eq.~(\ref{s16}).
Like spin squeezing, we neglect the decay of the macroscopic spin in the theoretical calculation, a decoherence process that can degrade light squeezing here.
To match the experimental results, we phenomenologically considering the effect of $T_1$, and get $\xi_L^2(T) = \left[ 1 - g_1 f_1(T) - g_2 f_2(T) \right] e^{2T/T_1}$, which is the fitting function for Fig.~2(a) in the main text.

Considering the quantitative difference between theory and experiment due to the assumptions made for analytical derivations, these derived functions are used to fit the trend of the experimental curves, with coefficients $b_{1,2}, c_{1,2}, d_{1,2}, e_{1,2}, g_{1,2}, h_{1,2}$ being free parameters in fitting.
The actual calculated squeezing is higher than the observed values because of these simplifications of the model.

\section{2. Experimental details}

\subsection{2.1. Experimental setup}

\begin{figure}[htbp]
\centering
\includegraphics[width=0.4\linewidth]{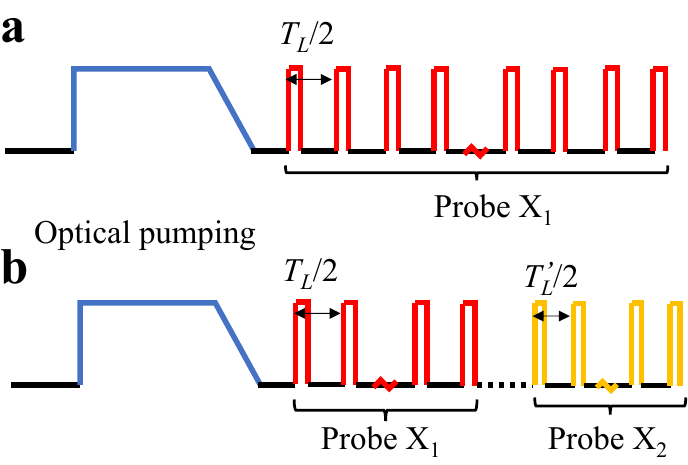}
\caption{Pulse sequence for (a) light squeezing and (b) spin squeezing.
$T_L = 2\pi/\Omega$ and $T_L' = 2\pi/\Omega'$ denote the periods of Larmor precession during the X$_1$- and X$_2$-probe, respectively.}
\label{figs3}
\end{figure}

The $7~\mathrm{mm}\times7~\mathrm{mm}\times20~\mathrm{mm}$ rectangular paraffin-coated rubidium vapor cell at $53.5^\circ\mathrm{C}$ is located in a four-layer magnetic shield.
A bias magnetic field of $0.71~\mathrm{G}$ is along $x$-axis. The optical pumping contains a pump laser resonant with the Rb D1 transition $5\mathrm{S}_{1/2}, F=2 \rightarrow 5\mathrm{P}_{1/2}, F'=2$ and a repump laser resonant with the Rb D2 transition $5\mathrm{S}_{1/2}, F=1 \rightarrow 5\mathrm{P}_{3/2}, F'=2$, sharing the same circular polarization and propagating along $x$-axis.
The probe laser $\rm{X}_1$ (X$_2$) is $1.66~\mathrm{GHz}$ red-detuned ($2.50~\mathrm{GHz}$ blue-detuned) from the Rb D2 transition $5\mathrm{S}_{1/2}, F=2 \rightarrow 5\mathrm{P}_{3/2}, F'=3$, with linear polarization along the $x$-($y$-)axis.
Both $\rm{X}_1$ and X$_2$ propagate along the $z$-axis, and their Stokes components $S_z$ and $S_y$ will be respectively demodulated at the Larmor frequency and recorded by a lock-in amplifier.

First, let's introduce the detection of light squeezing.
As shown in Fig.~\ref{figs3}(a), after $11~\mathrm{ms}$ optical pumping, a $1.0~\mathrm{ms}$ stroboscopic probe X$_1$ pulse is applied with stroboscopic frequency $\omega_m = 2\Omega$, duty cycle $d = 0.08$ and mean power $1.18~\mathrm{mW}$.
Both the Zeeman effect and AC Stark shift (induced by the strong probe light field) will lift the ground-state degeneracy, resulting in an effective Larmor frequency $\Omega = 2\pi \times 499.60~\mathrm{kHz}$ during the X$_1$ probe pulse.
The pumping and probing pulses both have slowly varying rising and falling edges \cite{PRJ.413288} of $100~\mathrm{\mu s}$, except for the falling edge of the X$_1$ pulse, which is $10~\mathrm{\mu s}$.
Because the squeezing component of light is $\hat{p}_L$, we use a quarter-wave plate (QWP) to make a balanced $S_z$ measurement on the $1.0~\mathrm{ms}$ X$_1$ signal pulse.
The measurement signal is sent to a lock-in amplifier with demodulation frequency $\Omega$ to output the in-phase and out-of-phase signals $\{X, Y\}$ , which is followed by a fast Fourier transform (FFT) on $X+iY$ to get the light spectrum.
We repeat this process $10,000$ times and take an average to get the LSS spectrum $S(\omega)_\mathrm{LSS}$.
We also detect the spectrum of SN $S(\omega)_\mathrm{SN}$ using the same experimental setup, but without the presence of the atomic ensemble.
In order to minimize the influence of technical noise, we exclude the data at frequency $\Omega$ in $S(\omega)_\mathrm{SN}$ and fit the rest data points with Lorentz function to acquire the shot noise level, then divide $S(\omega)_\mathrm{LSS}$ with fitted $S(\omega)_\mathrm{SN}$ to obtain the light squeezing $\xi_{L}^2(\omega) = S(\omega)_\mathrm{SL}/S(\omega)_\mathrm{SN}$.

Next, we describe the verification process of spin squeezing.
In order to verify the generated SSS at the end of X$_1$ pulse, we use a $2.50~\mathrm{GHz}$ blue-detuned far off-resonant laser, called probe X$_2$, with linear polarization along $y$ axis to perform a QND-type measurement of atoms \cite{NP23, Nature.581.7807}.
As shown in Fig.~\ref{figs3}(b), after $11~\mathrm{ms}$ of optical pumping, a $1.0~\mathrm{ms}$ stroboscopic probe X$_1$ pulse is applied to atoms, with the same parameters as in light squeezing.
Right after the X$_1$ pulse, a $13~\mathrm{ms}$ stroboscopic probe X$_2$ pulse is turned on with stroboscopic frequency $\omega_m' = 2\Omega' = 2\times 2\pi \times 499.93 \mathrm{kHz}$, duty cycle $d' = 0.1$ and mean power $1.24~\mathrm{mW}$.
The difference between the two stroboscopic frequencies, $\Omega$ and $\Omega'$, arises from the variation in detuning and laser power, which leads to a different AC Stark shift.
X$_2$ pulse is also amplitude modulated by slowly varying the rising and falling edges with $100~\mathrm{\mu s}$ length.
As in the typical QND measurement, we use a half-wave plate (HWP) to make a balanced $S_y$ measurement on X$_2$ pulse to reconstruct the squeezed spin component $\hat{p}_{A}$ from the detected $\hat{x}_{L}$ signal.
The degree of generated spin squeezing is determined by Wineland criterion \cite{PhysRevA.50.67} $\xi_{A,W}^2 = e^{2T/T_1}\mathrm{Var}(\hat{p}_A)_\mathrm{SSS}/\mathrm{Var}(\hat{p}_A)_\mathrm{PNL}$, where $T_1=18~\mathrm{ms}$ as mentioned above.
We also apply a decaying time mode $f(T') = e^{-\gamma' T'}$ to X$_2$ pulse with $\gamma'^{-1} = 1.3~\mathrm{ms}$ before performing the variance calculation to optimize the efficiency of atomic state reconstruction.
Such a time mode renders the data at longer time (here, beyond $2.6~\mathrm{ms}$) in $T'$ negligible.

To calibrate the PNL $\mathrm{Var}(\hat{p}_A)_\mathrm{PNL}$, or the noise of ideal CSS with $100\%$ polarization, we use the same method as in references \cite{NP23, Nature.581.7807}, and will introduce it briefly here.
Under the QND Hamiltonian, the atomic variance is detected by $S_y$ ($\hat{x}_{L}^\mathrm{out}$) component.
When the ensemble is in an ideal CSS, the variance of the transverse spin is $\mathrm{Var}(\hat{j}_y) = \mathrm{Var}(\hat{j}_z) = F/2 = 1$ (with $\langle j_x^2\rangle = F^2$).
When the ensemble is in thermal state, where the atoms are evenly spread among the ground states, the variance becomes $\mathrm{Var}(\hat{j}_y) = \mathrm{Var}(\hat{j}_y) = \mathrm{Var}(\hat{j}_z) = F(F+1)/3 = 2$.
Considering the fact that only 5/8 of total atoms are in the hyperfine state $|F=2\rangle$, the detected variance under thermal state is 1.25 times the PNL.
Letting the noise of $S_y$ measurement be TN under the thermal spin state, SSS under the squeezed spin state, SN the shot noise without atoms, then the squeezing degree can be calculated as $\xi_{A,W}^2 = 1.25\times e^{2T/T_1} (\mathrm{SSS}-\mathrm{SN})/(\mathrm{TN}-\mathrm{SN})$.

\subsection{2.2. Decoherence processes}

Here we consider the decay of the atoms, which is caused by the spontaneously emitted radiation and atomic collisions with atoms and cell walls.
There are two different types of decoherence processes in the experiment.
One is the extra decay of the transverse collective-spin components at rate $\gamma_{\rm{ex}}$ \cite{BrainPHD}, and the other is the macroscopic-spin decay with time $T_1$.
For $\gamma_{\rm{ex}}$, we can re-write the propagation equations [Eq.~(\ref{eq:whole4})] as:
\begin{subequations}
\label{eq:whole42}
\begin{eqnarray}
\hat x_L^{{\rm{out}}}(t) &=& \hat x_L^{{\rm{in}}}(t) + \kappa \phi (t)\left[ { - {{\hat x}_A}(t)\sin (\Omega t) + {{\hat p}_A}(t)\cos (\Omega t)} \right],\label{s11a2}\\
\hat p_L^{{\rm{out}}}(t) &=& \hat p_L^{{\rm{in}}}(t) - {\zeta ^2}\kappa \phi (t)\left[ {{{\hat x}_A}(t)\cos (\Omega t) + {{\hat p}_A}(t)\sin (\Omega t)} \right],\label{s11b2}\\
\partial_t{{\hat x}_A}(t) &=&  - {\gamma _s}{{\hat x}_A}(t) + \kappa \phi (t)\hat p_L^{{\rm{in}}}(t)\cos (\Omega t) + {\zeta ^2}\kappa \phi (t)\hat x_L^{{\rm{in}}}(t)\sin (\Omega t) - {\gamma _{{\rm{ex}}}}{{\hat x}_A}(t) + \sqrt {2{\gamma _{{\rm{ex}}}}} {{\hat f}_x}(t),\label{s11c2}\\
\partial_t{{\hat p}_A}(t) &=&  - {\gamma _s}{{\hat p}_A}(t) + \kappa \phi (t)\hat p_L^{{\rm{in}}}(t)\sin (\Omega t) - {\zeta ^2}\kappa \phi (t)\hat x_L^{{\rm{in}}}(t)\cos (\Omega t) - {\gamma _{{\rm{ex}}}}{{\hat p}_A}(t) + \sqrt {2{\gamma _{{\rm{ex}}}}} {{\hat f}_p}(t).\label{s11d2}
\end{eqnarray}
\end{subequations}
Here $\hat{f}_{x,p}$ are Langevin noise operators \cite{scully_quantum_1997} obeying $\langle \hat{f}_x(t) \hat{f}_x(t') \rangle = \langle \hat{f}_p(t) \hat{f}_p(t') \rangle = \delta(t-t')/2$, $\langle \hat{f}_x(t) \hat{f}_p(t') \rangle = 0$.
Following the similar process in the former section, we can derive the variance of $\hat p_A$ as:
\begin{eqnarray} \label{eqVt2}
\mathrm{Var}( \hat{p}_A)_\mathrm{SSS} &=& \frac{1}{2} \left[e^{-2\gamma T} +  \varepsilon\left(1-e^{-2\gamma T}\right) \mathcal{D}_{A} + \mathcal{F}_{A} \right],\\
\mathcal{F}_{A} &=& \left( 1 -\varepsilon \right)\left(1-e^{-2\gamma T}\right),\nonumber
\end{eqnarray}
with $\mathcal{D}_{A}$ the same as Eq.~(\ref{eqVt}) and $\gamma = \gamma_s + \gamma_\mathrm{ex}$, $\varepsilon = \gamma_s/\gamma$.
We can see from Eq. (\ref{eqVt}) and Eq. (\ref{eqVt2}) that the presence of atomic decay diminishes the maximum achievable spin squeezing, while simultaneously accelerating the rate of squeezing due to the larger value of $\gamma$. However, it should be stressed that the transverse decay will not change the trend of squeezing, that is, the degree of squeezing gradually increases and ultimately reaches the upper bound $\epsilon \mathcal{D}_A+\mathcal{F}_A$ (it is $\mathcal{D}_A$ for the ideal case).
Similarly, the light squeezing spectrum of $\hat{p}_{L}^\mathrm{out}$ becomes:
\begin{eqnarray}\label{eqSw2}
        S(\omega)_\mathrm{LSS} &=& \frac{1}{2} \left\{1 - \sum_{k=-\infty}^{\infty} \frac{{\left[ \mathcal{D}_{L,\mathrm{corr}}(\frac{k-1}{2}) - \mathcal{D}_{L,\mathrm{QBA}}(\frac{k-1}{2}) - \mathcal{F}_{L}(\frac{k-1}{2}) \right] }}{{1 + {{\left[ {\omega  - k\Omega } \right]}^2/{\gamma ^2}}}}\right\},\label{s162}\\
        \mathcal{D}_{L,\mathrm{corr}}(n) &=& \varepsilon f_2 \alpha(n),\nonumber\\
        \mathcal{D}_{L,\mathrm{QBA}}(n) &=& \varepsilon \zeta^2 f_1 \alpha(n) + \varepsilon^2 \left( f_2-f_1 \right) \bigg[  \frac{1}{2} \left( 1+\zeta^4 \right) \alpha(n) + \left( 1-\zeta^4 \right) \beta(n) \bigg],\nonumber\\
        \mathcal{F}_{L}(n) &=& \varepsilon \left( 1-\varepsilon \right) \zeta^2 \left( f_2-f_1 \right) \alpha(n),\nonumber
\end{eqnarray}
where $\alpha(n), \beta(n), f_1(T)$, and $f_2(T)$ are the same as Eq.~(\ref{eqSw}) except for the exchange from $\gamma_{s}$ to $\gamma$.
The presence of $\gamma_{\rm{ex}}$ again reduces the maximal light squeezing and decreases the time to achieve it, and the trend of squeezing is also unchanged.

In order to determine the total squeezing rate $\gamma$, which is conventionally called the transverse decay rate $T_2^{-1} = \gamma$, we detect the decay of the transverse spin signal in a separate experiment.
According to Eqs.~(\ref{s13}) and (\ref{s14}), one may directly calculate the mean-value evolution
\begin{eqnarray}\label{eq:mean}
    \left\langle {{{\hat x}_A}(t)} \right\rangle  = \left\langle {{{\hat x}_A}(0)} \right\rangle {e^{ - \gamma t}},\left\langle {{{\hat p}_A}(t)} \right\rangle  = \left\langle {{{\hat p}_A}(0)} \right\rangle {e^{ - \gamma t}},
  \end{eqnarray}
where we have used the fact that $\langle\hat\vartheta_L^{\rm{in}}(t)\rangle=\langle\hat f_\vartheta(t)\rangle=0$.
These equations indicate that the mean values of the atomic canonical variables all damped exponentially with $\gamma t$.
So, we apply a radio frequency (RF) field along $z$ axis with Larmor frequency $\Omega$ between optical pumping pulse and probe pulse X$_1$.
Then the generated transverse spin $\hat{p}_A(0)$ will decay at the rate of $\gamma$, and the acquired mean value of light signal is $\langle\hat{x}_{L}^\mathrm{out}\rangle = -d\zeta^2\kappa \langle \hat{p}_A(t)\rangle\sin(\Omega t) = -d\zeta^2\kappa \langle \hat{p}_A(0)\rangle {e^{ - \gamma t}}\sin(\Omega t)$.
With the lock-in amplifier, we can fit the demodulated signal and finally get the total transverse decay rate $\gamma$.

For $T_1$, which is defined by $\left\langle J_x(t) \right\rangle = J_x(0)e^{-t/T_1}$, as it is much longer than $\gamma^{-1}$, we ignore its effect in theoretical calculation.
However, in a metrology application, not only the variance of atomic spin, but also the amount of macroscopic-spin will affect the signal-to-noise ratio.
So, we apply the Wineland criterion \cite{PhysRevA.50.67} $\xi_{A,W}^2 = e^{2T/T_1}\mathrm{Var}(\hat{p}_A)_\mathrm{SSS}/\mathrm{Var}(\hat{p}_A)_\mathrm{PNL}$.
We can see that the finite value of $T_1$ will decrease the final spin squeezing.
For example, under our experimental parameters, the noise reduction $\xi_{A}^2 = \mathrm{Var}(\hat{p}_A)_\mathrm{SSS}/\mathrm{Var}(\hat{p}_A)_\mathrm{PNL}$ is $1.02~\mathrm{dB}$ for $T=0.8~\mathrm{ms}$, and the power broadened longitudinal relaxation time is measured to be $T_1=18~\mathrm{ms}$, then the decrease of squeezing parameter will be $10~\mathrm{log}_{10}(e^{2T/T_1}) = 0.39~\mathrm{dB}$, leading to a final squeezing of $0.63~\mathrm{dB}$ in the main text.

\begin{figure}[htbp]
\centering
\includegraphics[width=0.5\linewidth]{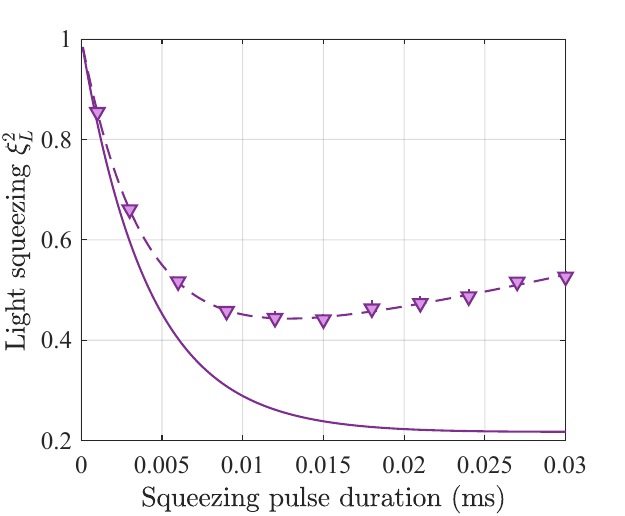}
\caption{Numerical simulation of light squeezing with and without the effect of $T_1$ (spin polarization decrease).
The triangle dots are simulation results with $T_1$ decay, and the solid line is the theory result without the consideration of $T_1$ decay.
We take $d=0.1$, $\kappa=143$, $\zeta^2=0.1$, $\gamma_\mathrm{ex}=2\pi\times 50~\mathrm{Hz}$, $\Omega=2\pi\times 500~\mathrm{kHz}$, then $\gamma_s=2\pi\times 1.08~\mathrm{kHz}$. We note that $T_2$ has only an effect on the maximal achievable squeezing, but does not change the dependence on the squeezing pulse duration. The error bar of the triangle symbol is the standard deviation from five independent simulation, which is smaller than the symbol.
The dashed line is to guide the eye.}
\label{figslsT1}
\end{figure}

In order to consider the effect of $T_1$ in light squeezing, we make a numerical simulation based on Eq.~(\ref{s11a2}), Eq.~(\ref{s11b2}) and
\begin{subequations}
\label{eq:whole42T1}
\begin{eqnarray}
\partial_t{{\hat x}_A}(t) &=&  - e^{-t/T_1}{\gamma _s}{{\hat x}_A}(t) + e^{-t/T_1}\kappa \phi (t)\hat p_L^{{\rm{in}}}(t)\cos (\Omega t) + e^{-t/T_1}{\zeta ^2}\kappa \phi (t)\hat x_L^{{\rm{in}}}(t)\sin (\Omega t) - {\gamma _{{\rm{ex}}}}{{\hat x}_A}(t) + \sqrt {2{\gamma _{{\rm{ex}}}}} {{\hat f}_x}(t),\nonumber\\
\partial_t{{\hat p}_A}(t) &=&  - e^{-t/T_1}{\gamma _s}{{\hat p}_A}(t) + e^{-t/T_1}\kappa \phi (t)\hat p_L^{{\rm{in}}}(t)\sin (\Omega t) - e^{-t/T_1}{\zeta ^2}\kappa \phi (t)\hat x_L^{{\rm{in}}}(t)\cos (\Omega t) - {\gamma _{{\rm{ex}}}}{{\hat p}_A}(t) + \sqrt {2{\gamma _{{\rm{ex}}}}} {{\hat f}_p}(t).\nonumber
\end{eqnarray}
\end{subequations}
The inserted $e^{-t/T_1}$ prefactor comes from the decay of the macroscopic spin $\left\langle J_x(t) \right\rangle$, which will gradually decrease the coupling strength $\kappa$ and thus the squeezing rate $\gamma_s$ at a long interaction time.
Then the simulated data of $\{\hat p_L^{{\rm{out}}}\}$ undergoes the fast Fourier transform (FFT) to get the light spectrum, which is compared with the spectrum from $\{\hat p_L^{{\rm{in}}}\}$ to get the light squeezing parameter, following the same process that will be introduced in the experiment.
Fig.~\ref{figslsT1} plots the performance of the scheme in the absence [solid line, Eq.~(\ref{s162})] and presence (dashed line with triangles) of
macroscopic-spin decay, which indicates that the macroscopic-spin decay seriously affects the light squeezing. Unlike the case of no macroscopic-spin decay, in which the light noises exhibit a monotone increase, spin depolarization results in an increase in (unwanted) light noise. The longer the squeezing time, the more (unwanted) light noise will emerge, leading to the presence of an optimal squeezing time. Such a theoretical result is consistent with the experiment result as shown in Fig.~2(a) in the main text.
So, we can conclude that the increase of light noise at a longer squeezing time mainly comes from the decrease of spin polarization.

In a separate experiment, we also measure the macroscopic-spin decay time $T_1$ (or the longitudinal decay time in tradition).
We inject a probe light X$_1$ into the cell along $x$ axis, opposite to the optical pumping light. At the same time, the probe light X$_1$ along $z$ axis is still kept on to ensure that the atomic system is identical to the one that produces squeezing.
The probe light along $x$ axis will experience the Faraday rotation, and its rotation angle is proportional to the polarization of the atomic ensemble.
Then, by pumping the atomic spin to sub-level $|F=2, m_F=-2\rangle$ and fitting the decay signal of the Faraday rotation, we can acquire the decay time $T_1$.

\subsection{2.3. Calibration of spin squeezing direction}

\begin{figure}[htbp]
    \begin{center}
    \includegraphics[width=0.5\linewidth]{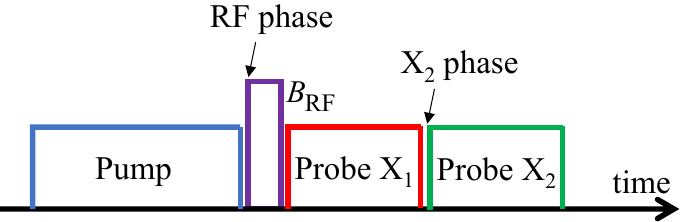}
    \caption{The pulse sequence for calibrating the measurement direction of the collective spin. $B_\mathrm{RF}$, RF field. X$_2$ phase, the initial phase of the probe X$_2$ pulse.}
    \label{sup_fig4}
    \end{center}
\end{figure}

In Fig.~3(b) of the main text, we detected spin squeezing at different directions.
However, because of the differences in the Larmor frequencies for the two probe processes (which is caused by the detuning and power differences of the two probe pulses as mentioned above) and the existence of the time gap between the two probe pulses (which will lead to a phase shift of X$_2$ pulse, see \cref{sup_fig4}) , it is necessary to calibrate the spin direction detected by X$_2$ pulse. To that end,
we first examine the mean-value evolution of the atomic operators caused by the X$_1$ pulse.
Eqs.~(\ref{eq:mean}) indicate that the mean values of the atomic canonical variables decay at the \emph{same rate}, that is, they are all damped exponentially with $\gamma t$.
Such a feature ensures that, for any input atomic coherent state,
 its direction will not be changed by the X$_1$ interaction.

Based on this feature, we developed a method for calibrating the spin direction, as shown in Fig.~\ref{sup_fig4}.
After the generation of CSS by the pumping pulse, we apply an RF field with frequency $\Omega$ (which is produced by a pair of transverse coils inside the magnetic shield) along the $z$ axis to the atomic ensemble to generate a nonzero transverse component, that is, $\langle\hat{p}_A\rangle\neq 0$.
After this procedure, the nonzero $\hat{p}_A$ component will precess around the $x$ axis at a frequency of $\Omega$ due to the presence of a magnetic field.
The phase of the RF field should be tuned to maximize the demodulated amplitude $R$ of the X$_1$ pulse, which ensures that, when the first time bin of the X$_1$ light enters the cell, the nonzero transverse component is exactly along the $z$ direction such that $\left\langle \hat{x}_A(0) \right\rangle = 0, \left\langle \hat{p}_A(0) \right\rangle \neq 0$.
As analyzed above, the X$_1$ interaction does not induce a rotation effect in the $\hat x_A$-$\hat p_A$ plane, but only slightly decreases the mean value of $\left\langle \hat{p}_A(0) \right\rangle$.
Therefore, the maximal mean value direction (in the phase space) after the X$_1$ interaction is still in the $\hat{p}_A$ direction.
To find out the $\hat{p}_A$ (or maximal-mean-value) direction, we also adjust the phase of the X$_2$ pulse.
Analogously, once the maximum demodulated amplitude $R$ of the X$_2$ pulse is detected, the phase will be fixed.
Such a phase ensures that the measurement direction of X$_2$ would be $\hat{p}_A$, corresponding to the rotation angle $\alpha = 0$ in the main text.
Other measurement directions can then be calibrated by simply adding a relative phase to the fixed phase of X$_2$.

\subsection{2.4. Discussion on experimental results}
\subsubsection{2.4.1. Dependency of concurrent squeezing on duty cycle}

The dependency of the concurrent squeezing on duty cycle, which is shown in Fig. 4 in the main text, can be understood by the physical picture introduced in the main text.
Let's considering the non-stroboscopic case ($d=1$) of the Eq. (\ref{eq1}): $\hat{H}_\mathrm{int} \propto \hat{b}^\dagger \hat\Gamma_{+}+ \mathrm{h.c.}$, indicating that the atomic mode interacts with the Bogoliubov mode $\hat\Gamma_{+}$ ($\hat\Gamma_{\pm}\equiv\mu_+ \hat{a}_{\pm 1} - \mu_- \hat{a}_{\mp 1}^\dag$, which can be produced by the \emph{two-mode-squeezing} transformation \cite{RevModPhys.77.513}) via the BS interaction.
Since the photonic mode $\hat\Gamma_{+}$ alone exhibits anti-squeezing \cite{muschik_efficient_2006}, no spin squeezing occurs after the BS (state-swapping) interaction;
for light modes, because they are in the two-mode squeezed state, light squeezing can be observed by detecting a linear combination of $\hat\Gamma_{+}$ and $\hat\Gamma_{-}$  \cite{RevModPhys.77.513}.
When the stroboscopic interaction takes place ($d<1$), the photonic single-mode squeezing appears according to interaction (\ref{eq1}).
With the decrease of $d$ (resulting in larger $r_k$), the photonic single-mode squeezing becomes larger, which contributes to both the enhancement of light squeezing and the creation of spin squeezing.

\subsubsection{2.4.2. Entanglement in the system}

First of all, squeezed spin state and squeezed light are both entanglement states, as described in the literature \cite{messikh_spin_2003, korbicz_spin_2005, toth_spin_2009}, especially, spin squeezed state is a spin-spin multipartite entanglement state.
Therefore, our demonstration of spin squeezing is equivalent to the demonstration of atomic entanglement state.
For light, we have demonstrated that two-mode squeezed light exists even without stroboscopic interaction ($d=1$) in Fig.~\ref{figs2}, which originates from the entanglement between the Stokes and anti-Stokes photon pair, i.e., the first up and lower sidebands photons.

However, about the entanglement between the vacuum light state (the Stokes and anti-Stokes modes) and the atomic mode, the short answer is no.
The role of the light and spin in our system is analogous to the two quantum fields in a parametric down conversion process.
It is known that \cite{ou_realization_1992, vahlbruch_detection_2016}, in the parametric down conversion process, when the two generated optical fields are entangled, then each of them by itself is noisier than a coherent state (such as super-possonian distribution); if each of them is squeezed, then they are not entangled with each other.
Here the spin and light are both squeezed, but they are barely entangled. We have theoretically calculated this and proved it is the case.

An independent theoretical investigation \cite{theoryWang} indicates that, a genuinely multipartite entangled state \cite{braunstein_quantum_2005} containing light modes and spin mode can be created, by appropriately modifying the Hamiltonian of Eq.~(\ref{eq1}).
However, then, we found that there is no spin squeezing. Because of the big difference in parameter regimes required to prepare such multipartite entanglement, this topic lies beyond the scope of this paper.

\subsubsection{2.4.3. Scaling of squeezing with optical depth}

The observed squeezing is relatively small mainly due to the hardware constraints in our experiment setup, in particular the geometry of the vapor cell, the property of the paraffin coating we have on the cell’s inner wall, and the available laser power. In the future, if we were to have a vapor cell of longer length, narrower cross section (so the total power is smaller for the same laser intensity), and paraffin coating that can endure higher temperature, then both the optical depth and the laser intensity would be higher, and we expect to obtain higher squeezing.

\begin{figure}[t]
\centering
\includegraphics[width=0.8\linewidth]{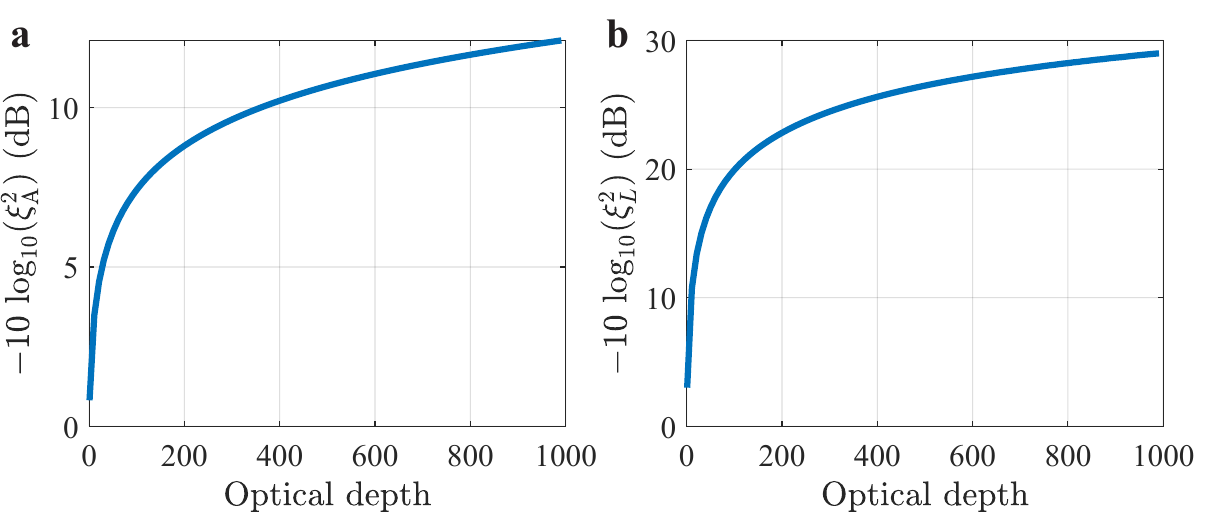}
\caption{The calculated (a) spin squeezing and (b) light squeezing versus optical depth of our proposed method. We note that strong squeezing can in principal be achieved with large optical depth.}
\label{figsod}
\end{figure}

In order to consider the scaling with optical depth, here we only consider the decoherence rate from spontaneous decay, then $\gamma _\mathrm{ex} = \eta /T$ and the optical depth $\alpha_0 = \kappa^2 / \eta$ \cite{hammerer_quantum_2010, theoryWang}, with $T$ the total squeezing time and $\eta$ the coefficient proportional to the spontaneous decay.
Similarly, we neglect the effect of $T_1$, and the amount of spin squeezing can be described by Eq.~(\ref{eqVt2}).
This expression shows that the spin squeezing parameter reaches its minimal at the limit $T \rightarrow \infty$ and $d \rightarrow 0$, then the squeezing parameter can be simplified to
\begin{equation}
\xi _{A, \mathrm{min}} ^2 = \varepsilon\zeta^2 + 1-\varepsilon = \frac{1 + \alpha_0\zeta^4}{1 + \alpha_0\zeta^2}.
\end{equation}
We can see that for a certain optical depth $\alpha_0$, there exist an optimal $\zeta^2 = (\sqrt{1+\alpha_0}-1)/\alpha_0$ to achieve the minimal squeezing parameter $\xi _A^2 = 2(\sqrt{1+\alpha_0}-1)/\alpha_0$.
The scaling of the optimal spin squeezing with optical depth is shown in Fig.~\ref{figsod}(a).

For light squeezing, we expected to do the similar calculation to get its scaling with optical depth.
However, from the expression of light squeezing parameter [Eq.~(\ref{eqSw2})], obtaining the optimal squeezing time $T$ requires solving a transcendental equation, which is unnecessary in our discussion, so instead we make a numerical calculation to get the light squeezing at a given optical depth $\alpha_0$ with the optimal $T, d, \zeta$, and the results are shown in Fig.~\ref{figsod}(b).

In general, there exists several factors that limit the achievable amount of squeezing. For atoms, in order to increase the contribution of the tensor part of the Hamiltonian [the term proportional to $a_2$ in Eq.~(\ref{eq_Hint})], we reduced the detuning between light and atoms (in contrast to the QND interaction which is under a more off-resonant condition), which on the one hand enhances the strength of the desired coherent interaction but on the other hand also increases the unwanted spontaneous decay.
For instance, the population relaxation time $T_1$ is reduced from 50 ms (QND case) to 18 ms, while the transverse spin relaxation time $T_2$ is reduced from 13 ms (QND case) to 3 ms.
As we showed in the main text, the short $T_1$ leads to a degradation of the spin squeezing by about 0.39 dB, while the short $T_2$ will also increase the noise of the squeezed quadratures and thus degrade the amount of spin squeezing.
For light, it is much sensitive to the reflection losses (which are mainly caused by cell walls and photon detectors) \cite{hammerer_quantum_2010}.
Regardless of the spin squeezing, we showed light squeezing about 1.51 dB is achieved, which corresponds to a squeezing of 2.25 dB at the cell output when corrections for cell-wall reflection losses (about $12\%$ in total) are made.

In summary, we would like to emphasize that the moderate squeezing created by our experiment is not due to the limitation of the principal of our proposal, but mostly due to the large spin decoherence rate and light losses.
These limitations can be overcome by using a vapor cell with larger optical depth (or with a cavity) and with improved antireflection coating, as shown in Fig.~\ref{figsod}.

\section{3. Proposed applications of currently squeezed light and spin}

In comparison with producing squeezed light and squeezed spin separately in different setups, the advantage of simultaneously squeezing them in one setup is that, the squeezed light is mode matched to the atoms, especially in the frequency spectrum. This allows further interaction of the squeezed light with the atoms, and especially when the atoms are already squeezed as in our protocol, interesting and useful results can appear for applications in quantum metrology and quantum networks. Here we propose three protocols to show how the dual squeezing could be taken advantage of. It is expected that more applications will emerge in the future.

\subsection{3.1. Enhancing the sensitivity of quantum metrology}

\begin{figure}[htbp]
    \begin{center}
    \includegraphics[width=0.5\linewidth]{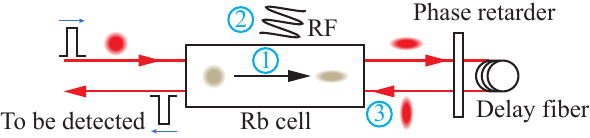}
    \caption{Proposed setup for measurement of an RF field amplitude using the dual squeezed state. A light pulse is sent through the vapor cell, squeezed, and then directed into a fiber optical delay line. While the pulse is in the fiber, an RF is applied to the atoms. After the application of RF, the light pulse is reflected back into cell again. A detection of the output light pulse will acquire the information about the RF field.}
    \label{sup_fig5}
    \end{center}
\end{figure}

Once the concurrent spin squeezing and light squeezing is generated, both squeezing may be used to improve the measurement accuracy. For example, let us consider the setup in Fig.~\ref{sup_fig5}. First, a coherent light pulse travels through the rubidium sample prepared in the coherent spin state and undergoes the interaction discussed in the main text. Here, for simplicity we assume the ideal stroboscopic limit ($d\rightarrow 0$), and then the input-output relations of Eq.~(\ref{eq:whole4}) can be expressed as
\begin{subequations}
\label{eq:whole42T2}
\begin{eqnarray}
\hat x_L^{{\rm{out1}}}(t) &=& \hat x_L^{{\rm{in}}}(t) + \kappa {\hat p}_A(t),\\\label{s27a}
\hat p_L^{{\rm{out1}}}(t) &=& \hat p_L^{{\rm{in}}}(t) - {\zeta ^2}\kappa  {\hat x}_A(t),\\
\hat{x}_A(t) &=& \hat{x}_A(0)e^{-\gamma_s t} + \kappa\int_{0}^{t}\mathrm{d}\tau~ e^{-\gamma_s(t-\tau)} \hat{p}_L^\mathrm{in}(\tau),\\
\hat{p}_A(t) &=&\hat{p}_A(0)e^{-\gamma_s t} - \zeta^2\kappa\int_{0}^{t}\mathrm{d}\tau~ e^{-\gamma_s(t-\tau)}\hat{x}_L^\mathrm{in}(\tau).
\end{eqnarray}
\end{subequations}
Next, we define the collective light mode
\begin{eqnarray*}
\hat \vartheta _{L \pm }^{{\rm{in/out1}}} = \sqrt {\frac{{2{\gamma _s}}}{{ \pm \left( {{e^{ \pm 2{\gamma _s}T}} - 1} \right)}}} \int_0^T {\rm{d}} \tau \;{e^{ \pm {\gamma _s}\tau }}\hat \vartheta _{L \pm }^{{\rm{in/out1}}}(\tau ),~\hat \vartheta  \in \left\{ {\hat x,\hat p} \right\}.
\end{eqnarray*}
Then we can derive the input-output relations for light and atoms from Eq. (\ref{eq:whole42T2}), obtaining
\begin{eqnarray}
\hat x_{L - }^{{\rm{out1}}} &=& \frac{1}{\zeta }\sqrt {1 - {e^{ - 2{\gamma _s}T}}} \hat p_A^{\rm{in}} + {e^{ - {\gamma _s}T}}\hat x_{L + }^{{\rm{in}}},
\hat p_{L - }^{{\rm{out1}}} =  - \zeta \sqrt {1 - {e^{ - 2{\gamma _s}T}}} \hat x_A^{\rm{in}} + {e^{ - {\gamma _s}T}}\hat p_{L + }^{{\rm{in}}},\label{s28}\\
\hat x_A^{{\rm{out1}}} &=& \hat x_A^{{\rm{in}}}{e^{ - {\gamma _s}T}} + \frac{1}{\zeta }\sqrt {1 - {e^{ - 2{\gamma _s}T}}} \hat p_{L + }^{{\rm{in}}},
\hat p_A^{{\rm{out1}}} = \hat p_A^{{\rm{in}}}{e^{ - {\gamma _s}T}} - \zeta \sqrt {1 - {e^{ - 2{\gamma _s}T}}} \hat x_{L + }^{{\rm{in}}},\label{s29}
\end{eqnarray}
where $\hat x_A^{\rm{in}}=\hat x_A(0),\hat p_A^{\rm{in}}=\hat p_A(0)$, $\hat x_A^{\rm{out1}}=\hat x_A(T),\hat p_A^{\rm{out1}}=\hat p_A(T)$ denote the quadratures of the atomic state before, and after interactions. According to the results outlined above, the $\hat p$ quadrature of both the atomic spin and the output light will be squeezed. This can be seen immediately from Eqs. (\ref{s28}) and (\ref{s29}) in the limit $\gamma_s T\rightarrow \infty$, resulting in $\hat p_{L - }^{{\rm{out1}}} =  - \zeta \hat x_A^\mathrm{in},\hat p_A^{{\rm{out1}}} =  - \zeta \hat x_{L + }^{{\rm{in}}} \Rightarrow \text{Var}(\hat p_{L - }^{{\rm{out1}}}) = \text{Var}(\hat p_A^{{\rm{out1}}}) = {\zeta ^2}/2<1/2$.

 After the interaction, an RF field with the amplitude $B_\mathrm{RF}$ to be measured is applied to the atomic ensemble, resulting in a displacement of $\hat{p}_A^{\rm{out1}}$, yielding $\hat{x}_A^{\rm{out2}}=\hat{x}_A^{\rm{out1}},\hat{p}_A^{\rm{out2}}=\hat{p}_A^{\rm{out1}}+p_0$ with $p_0 \propto B_\mathrm{RF}$. Next, after passing through a delay fiber and a phase retarder, the output light is reflected back into the cell again to experience a second interaction. Here, the delay fiber promise enough time for the interaction between the atoms and the application of the RF field, and the double-pass through the phase retarder rotates the Stokes operator by $90^\circ$, which changes the squeezing direction of light from $\hat{p}_L$ to $\hat{x}_L$, resulting in $\hat x_{L + }^{{\rm{out2}}} = \hat p_{L + }^{{\rm{out1}}},\hat p_{L + }^{{\rm{out2}}} =  - \hat x_{L + }^{{\rm{out1}}}$. Finally, once the second interaction is completed, we obtain the output position quadrature $\hat x_L$ in the following form:
\begin{eqnarray}
\hat x_{L - }^{{\rm{out3}}} &=& \frac{1}{\zeta }\sqrt {1 - {e^{ - 2{\gamma _s}T}}} \hat p_A^{{\rm{out}}2} + {e^{ - {\gamma _s}T}}\hat x_{L + }^{{\rm{out2}}}\nonumber\\
 &=& \frac{1}{\zeta }\sqrt {1 - {e^{ - 2{\gamma _s}T}}} \left( {\hat p_A^{{\rm{in}}}{e^{ - {\gamma _s}T}} - \zeta \sqrt {1 - {e^{ - 2{\gamma _s}T}}} \hat x_{L + }^{{\rm{in}}}} \right) + {e^{ - {\gamma _s}T}}\hat p_{L+}^{\rm{out1}}\nonumber\\
 &&+ \frac{1}{\zeta }\sqrt {1 - {e^{ - 2{\gamma _s}T}}} {p_0}.\label{s30}
\end{eqnarray}
Obviously, information about the RF field can be extracted by detecting $\hat x_{L - }^{{\rm{out3}}}$.
Here the last term $p_0$ denotes the signal (proportional to $B_{RF}$) while the rest terms represent the (unwanted) noises, which blurs the information about the signal and should be suppressed. The pre-squeezing (produced by the first interaction) does contribute to suppress such noises. For atoms, due to the spin squeezing the variance in the bracket become $[{e^{ - 2{\gamma _s}T}}(1 - {\zeta ^2}) + {\zeta ^2}]/2<1/2$; for light, due to the light squeezing the variance of $\hat p_{L +}^{{\rm{out1}}}$ is $[1 - 4\gamma _s^2{T^2}(1 - {\zeta ^2})/({e^{2{\gamma _s}T}} - 1)]/2<1/2$.
Therefore, both pre-squeezing would benefit the measurement accuracy.
 From Eq. (\ref{s30}) one may directly calculate the signal to noise ratio (SNR), obtaining
\begin{eqnarray}
\mathrm{SNR} = \frac{\sqrt{2} p_0}{\sqrt{\zeta^2 + (1-\zeta^2) e^{-2\gamma_s T} + \frac{\zeta^2}{e^{2\gamma_s T} - 1}\Big[1-\frac{4\gamma_s^2 T^2 (1-\zeta^2)}{e^{2\gamma_s T} - 1}\Big]}} \to \frac{{\sqrt 2 p_0}}{\zeta },~~~{\gamma _s}T \gg 1,
\end{eqnarray}
indicating that large SNRs are obtainable for small values of $\zeta$. For comparison, let us consider the RF magnetometry based on the traditional QND interaction $\hat{H}_\mathrm{QND} = \kappa \hat{p}_L \hat{p}_A$. First, the atomic ensemble is displaced, resulting in the input-output relations with the displacement $p_0 \propto B_\mathrm{RF}$ as $\hat{p}_A^\mathrm{out} = \hat{p}_A^\mathrm{in} + p_0$. Next, the atomic state is read out via $\hat{H}_\mathrm{QND}$ interaction, yielding  $\hat{x}_L^\mathrm{out} = \hat{x}_L^\mathrm{in} + \kappa \hat{p}_A^\mathrm{out} = \hat{x}_L^\mathrm{in} + \kappa \hat{p}_A^\mathrm{in} + \kappa p_0$. One can easily derive the SNR from this relation and gives $\mathrm{SMR}_\mathrm{QND} = \sqrt{2} \kappa p_0 / \sqrt{1+\kappa^2}$. This SNR, however, can only reach $\sqrt{2} p_0$ even for infinite coupling strength $\kappa$, and, in fact, the limitation mainly comes from the intrinsic spin projection noise limit. Therefore, the pre-squeezing can increase the SNR by a factor of $1/\zeta$.

Compared to traditional schemes where the squeezed light is generated externally and then injected into the atomic medium, our dual-squeezing protocol ensures that the squeezed light has a frequency and spectrum that matches the atomic transitions, and in particular the squeezed light is already in a stroboscopic form which is required by the backaction-evading measurement. In addition, experimental resources are saved by omiting a separate setup dedicated to producing squeezed light.

\subsection{3.2. Improving spin squeezing by injected squeezed light}
The generated light squeezing in our protocol can also be used to further improve the already produced spin squeezing. Here, we would consider a double-pass scheme. The setup is similar to Fig.~\ref{sup_fig5}, but with the RF signal absent. Then, after the first interaction, the squeezed light pulse is directly reflected back into the atomic ensemble again to experience a second interaction described by Eq. (\ref{eqs7}), yielding the input-output relations for atomic momentum quadrature
\begin{eqnarray}
\hat p_A^{{\rm{out2}}} &=& {e^{ - {\gamma _s}T}}\hat p_A^{{\rm{out1}}} - \zeta \sqrt {1 - {e^{ - 2{\gamma _s}T}}} \hat p_{L + }^{{\rm{out1}}},\\
 \Rightarrow {\mathop{\rm var}} \left( {\hat p_A^{{\rm{out2}}}} \right) &=& \frac{1}{2}\left[ {{e^{ - 4{\gamma _s}T}} + {\zeta ^2}\left( {1 - {e^{ - 4{\gamma _s}T}}} \right)} \right] - 2\gamma _s^2{T^2}{\zeta ^2}(1 - {\zeta ^2}){e^{ - 2{\gamma _s}T}}.\label{s33}
\end{eqnarray}
Here, comparing to the amount of squeezing for single pass, that is, ${\mathop{\rm var}} ( {\hat p_A^{{\rm{out1}}}} ) = \frac{1}{2}[ {{e^{ - 2{\gamma _s}T}} + {\zeta ^2}( {1 - {e^{ - 2{\gamma _s}T}}} )} ]$, we can see that the first part (in square bracket) of Eq. (\ref{s33}) is just replaced by $\gamma_s\rightarrow 2\gamma_s$, which is the traditional result when the optical depth of the atomic system is effectively doubled.
However, in our double-pass scheme, with the help of the squeezed light generated from the first interaction, the final spin squeezing is more than that just doubling the optical depth, which is indicated by the last negative term of Eq. (\ref{s33}).
Therefore, our double-pass scheme with the dual squeezing can achieve higher spin squeezing than just doubling the length of the cell.

\subsection{3.3. Entangling quantum memory nodes in a quantum network}

\begin{figure}[htbp]
    \begin{center}
    \includegraphics[width=1.0\linewidth]{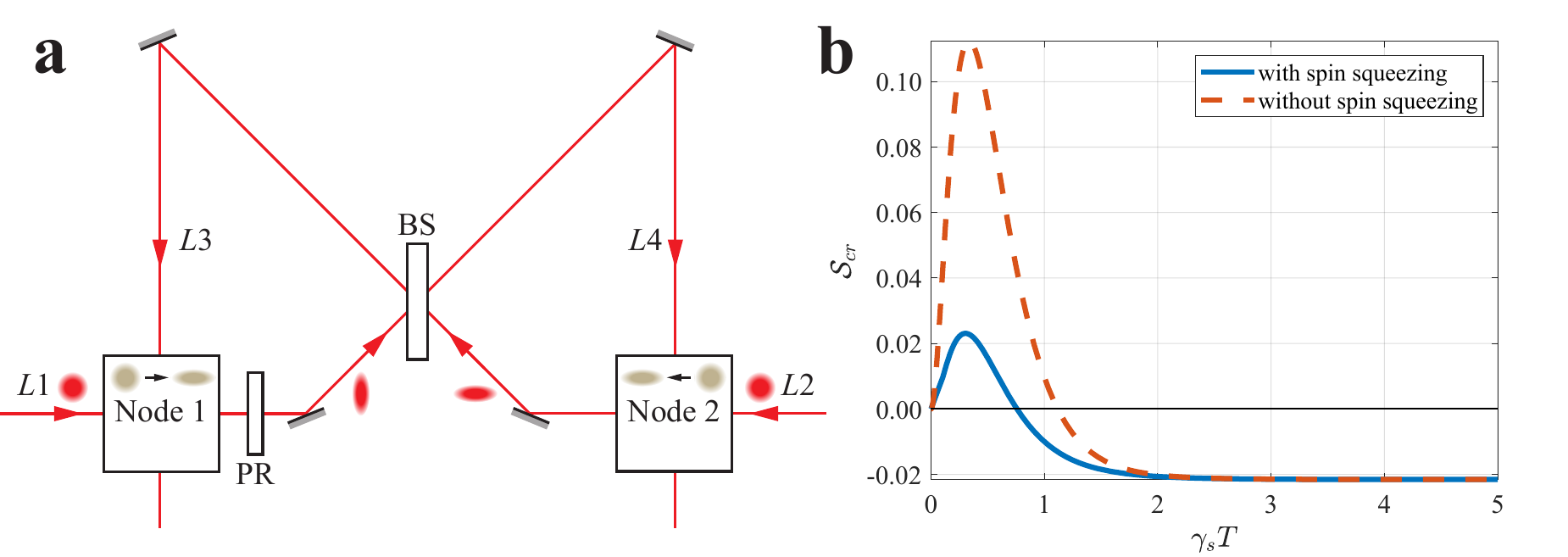}
    \caption{(a) Proposed setup for entangling quantum memory nodes using the dual squeezed state. Two light pulses are sent through the vapor cells, which will be squeezed, and then mixed at a beam splitter to generate two-mode squeezed states of light. Then the light pulses are reflected back into the nodes again, resulting in the entanglement of the two pre-squeezed nodes. PR, phase retarder; BS, beam splitter.
    (b) Separability versus $\gamma_s T$ with and without spin squeezing. $\mathcal{S}_{cr} < 0$ indicates entanglement between two nodes. $\mathrm{var}(\hat{p}_{A1,A2}^\mathrm{out1}) = \mathrm{var}(\hat{x}_{A1,A2}^\mathrm{out1}) = 1$ without spin squeezing; $\mathrm{var}(\hat{p}_{A1,A2}^\mathrm{out1}) = e^{-0.6}$, $\mathrm{var}(\hat{x}_{A1,A2}^\mathrm{out1}) = e^{0.6}$ with spin squeezing. Here we have taken $r=0.5$ and $ \zeta = 0.2$.}
    \label{sup_fig9}
    \end{center}
\end{figure}

Atomic ensembles can be used for quantum memory for light and are commonly referred to as the quantum memory nodes in a quantum network \cite{kimble_quantum_2008}. Establishing remote entanglement between two or more quantum memory nodes is an important element for the realization of quantum network, which enables applications such as quantum repeater \cite{azuma_quantum_2023}, distributed quantum computation \cite{van_meter_distributed_2010}, and distributed quantum metrology \cite{ge_distributed_2018}. Here we show that entanglement between remote atomic ensembles can be generated by utilizing the concurrent spin and light squeezing. Considering a setup shown in Fig.~\ref{sup_fig9}(a), two atomic ensembles, labeled $1$ and $2$, are located at a considerable distance from each other. To establish the entanglement between the two atomic ensembles, two light pulses (also labeled $1$ and $2$) are simultaneously sent through the atoms to create concurrent spin and light squeezing. After propagating some distance, the pulse $1$ first enters a phase retarder to rotate the quadratures $\hat x_{L1}\rightarrow \hat p_{L1},\hat p_{L1}\rightarrow -\hat x_{L1}$, and then the two pulses are combined on a balanced beam splitter. Since the light pulse $1$ before entering the BS is squeezed in $\hat x$ while the light pulse $2$ is squeezed in $\hat p$, the two light pulses, labeled $3$ and $4$, at the two output ports are in the two-mode squeezed state \cite{RevModPhys.77.513}, or, in other words, they get entangled. Next, the two entangled pulses $3$ and $4$ are reflected back into the atomic ensembles $1$ and $2$, respectively, to experience a second interaction described by Eq. (\ref{eqs7}), leading to the input-output relations for the two atomic ensembles
\begin{eqnarray}
\hat x_{A1}^{{\rm{out2}}} &=& \hat x_{A1}^{{\rm{out1}}}{e^{ - {\gamma _s}T}} + \frac{1}{\zeta }\sqrt {1 - {e^{ - 2{\gamma _s}T}}} \hat p_{L3 + }^{{\rm{in}}},\hat p_{A1}^{{\rm{out2}}} = \hat p_{A1}^{{\rm{out1}}}{e^{ - {\gamma _s}T}} - \zeta \sqrt {1 - {e^{ - 2{\gamma _s}T}}} \hat x_{L3 + }^{{\rm{in}}},\label{eqs34}\\
\hat x_{A2}^{{\rm{out2}}} &=& \hat x_{A2}^{{\rm{out1}}}{e^{ - {\gamma _s}T}} + \frac{1}{\zeta }\sqrt {1 - {e^{ - 2{\gamma _s}T}}} \hat p_{L4 + }^{{\rm{in}}},\hat p_{A2}^{{\rm{out1}}} = \hat p_{A2}^{{\rm{out1}}}{e^{ - {\gamma _s}T}} - \zeta \sqrt {1 - {e^{ - 2{\gamma _s}T}}} \hat x_{L4 + }^{{\rm{in}}},\label{eqs35}
\end{eqnarray}
where $\hat \vartheta ^{\rm{in}}_{3+}$ and $\hat \vartheta ^{\rm{in}}_{4+}$ denote the light quadratures of light modes $3$ and $4$, respectively. From these equations, one may derive the correlation matrix of the bipartite two-mode system in block form \cite{RevModPhys.77.513},
\begin{eqnarray}
{V^{\left( 2 \right)}} = \left( {\begin{array}{*{20}{c}}
A&C\\
{{C^T}}&B
\end{array}} \right),\label{eqs36}
\end{eqnarray}
where
\begin{eqnarray}
A &=& \frac{1}{4}\left( {\begin{array}{*{20}{c}}
{{\mathop{\rm var}} \left( {\hat x_{A1}^{{\rm{out1}}}} \right){e^{ - 2{\gamma _s}T}} + \frac{{1 - {e^{ - 2{\gamma _s}T}}}}{{{\zeta ^2}}}\cosh 2r}&0\\
0&{{\mathop{\rm var}} \left( {\hat p_{A1}^{{\rm{out1}}}} \right){e^{ - 2{\gamma _s}T}} + {\zeta ^2}\left( {1 - {e^{ - 2{\gamma _s}T}}} \right)\cosh 2r}
\end{array}} \right),\\
B &=& \frac{1}{4}\left( {\begin{array}{*{20}{c}}
{{\mathop{\rm var}} \left( {\hat x_{A2}^{{\rm{out1}}}} \right){e^{ - 2{\gamma _s}T}} + \frac{{1 - {e^{ - 2{\gamma _s}T}}}}{{{\zeta ^2}}}\cosh 2r}&0\\
0&{{\mathop{\rm var}} \left( {\hat p_{A2}^{{\rm{out1}}}} \right){e^{ - 2{\gamma _s}T}} + {\zeta ^2}\left( {1 - {e^{ - 2{\gamma _s}T}}} \right)\cosh 2r}
\end{array}} \right),\\
C &=& \frac{1}{4}\left( {\begin{array}{*{20}{c}}
{ - \frac{{1 - {e^{ - 2{\gamma _s}T}}}}{{{\zeta ^2}}}\sinh 2r}&0\\
0&{{\zeta ^2}\left( {1 - {e^{ - 2{\gamma _s}T}}} \right)\sinh 2r}
\end{array}} \right).
\end{eqnarray}
In deriving Eq. (\ref{eqs36}), we have assumed $\text{var}(\hat x_{L3+}^{\rm{in}}-\hat x_{L4+}^{\rm{in}})=\text{var}(\hat p_{L3+}^{\rm{in}}+\hat p_{L4+}^{\rm{in}})=e^{-2r}$ with $r$ being the two-mode squeezing parameter \cite{RevModPhys.77.513}. To find whether there exist entanglement between the two atomic ensembles, we use the Simon's separability criterion \cite{simon_peres-horodecki_2000}
\begin{eqnarray}
\mathcal{S}_{cr}=\det A\det B + {\left( {\frac{1}{{16}} - |\det C|} \right)^2} - {\mathop{\rm Tr}\nolimits} \left( {AJCJBJ{C^T}J} \right) - \frac{1}{{16}}\left(\det A + \det B\right)\geq 0. \label{eqs40}
\end{eqnarray}
Any separable bipartite state satisfies the inequality of (\ref{eqs40}), and therefore the violation of this inequality announces the presence of entanglement between the two atomic ensembles. Fig.~\ref{sup_fig9}(b) plots $\mathcal{S}_{cr}$ as a function of $\gamma T$, demonstrating that  (i) entanglement between atomic ensembles can be created by using the squeezed light generated, and (ii) spin squeezing can accelerate the process of entanglement generation. Thus, the concurrent generation of spin and light squeezing is beneficial for the remote entanglement creation between quantum memory nodes. This protocol can be generalized to entangling more than two atomic ensembles, by using more beam splitters as in standard multi-photon entanglement schemes~\cite{wang_experimental_2016, yan_establishing_2017}.

\bibliography{sref}